\documentclass[journal=aaemcq,manuscript=article]{achemso}
\setkeys{acs}{doi=true, maxauthors=99}

\usepackage{graphicx}
\usepackage{booktabs}
\usepackage{multirow}
\usepackage{amsmath}
\usepackage{amssymb}
\usepackage{subcaption}
\usepackage{tikz}
\usepackage{placeins}
\usepackage{xcolor}
\usetikzlibrary{arrows.meta,positioning,fit,backgrounds}
\graphicspath{{Figures/}}

\newcommand{\feat}[1]{\texttt{#1}}
\newcommand{\EgPBE}{E_g^{\mathrm{PBE}}}
\newcommand{\EgHSE}{E_g^{\mathrm{HSE06}}}
\newcommand{\dEg}{\Delta E_g}

 \title{Repairing PBE-Spurious Metallicity for HSE06-Level Screening of 2D Photocatalysts for Green Hydrogen Production}

\author{Ritam Chakraborty}
\affiliation{Theoretical Sciences Unit, Jawaharlal Nehru Centre for Advanced Scientific Research (JNCASR), Bangalore 560064, India}
\email{ritamchakraborty454@gmail.com}
\author{Arpan Das}
\affiliation{Department of Science and Humanities, Audisankara (Deemed to be University), Gudur 524101, India}
\email{arpandas9236@gmail.com}

\abbreviations{PBE,HSE06,C2DB,DFT,RF,GB,SVC,SVR,XGB,MAE,CV,CBM,VBM}
\keywords{2D materials, C2DB, $\Delta$-learning, HSE06 band gap, machine learning, target leakage, photocatalytic water splitting, solar hydrogen}

\begin{document}

\begin{abstract}
Semilocal PBE calculations can remove viable photocatalysts before screening by labeling narrow-gap semiconductors as metals. We address this failure mode in the Computational 2D Materials Database (C2DB) by combining leakage-aware repair of PBE-spurious metallicity with HSE06--PBE $\Delta$-learning. Stage~I classifies HSE06-unknown PBE metals using structural, chemical, magnetic, and stability descriptors, while excluding HSE06/GW quantities and PBE electronic shortcuts. Stage~II learns $\EgHSE-\EgPBE$ for the corrected insulating population. The curated XGBoost regressor reconstructs HSE06 gaps with a mean absolute error of 0.108~eV ($R^2=0.989$), compared with 1.036~eV for raw PBE. The Stage~I classifier is used only for triage because the labeled true-metal class contains 29 materials; its best holdout performance gives 87.5\% accuracy, 0.286 true-metal recall, and 0.643 balanced accuracy. The corrected pH~0 electronic screen yields 10 strict and 29 initial relaxed green-hydrogen photocatalyst candidates. Four strict and 18 relaxed candidates would fail the same 1.6--2.8~eV gap window at the PBE level. Targeted VASP HSE06 calculations for six ML-predicted compounds give material-level MAEs of 0.417, 0.182, and 0.110~eV for the C2DB-native, Magpie+structural, and curated models, respectively; \feat{1AgBr-1} shifts above the upper gap cutoff, leaving 28 retained relaxed candidates. The workflow shows that high-fidelity correction must be evaluated by candidate membership, not only by global regression error.
\end{abstract}

\section{Introduction}
Photocatalytic water splitting is a materials route to green hydrogen because it converts solar energy directly into chemical fuel. A useful screening candidate must satisfy electronic constraints before more expensive calculations are justified: the gap must be large enough for overall water splitting but still compatible with visible-light absorption, and the band edges must straddle the water redox potentials on an absolute energy scale.\cite{wang2020domen,mai2022chemrev} Stability, optical absorption, carrier transport, excitonic effects, and catalytic overpotentials remain essential later filters, but the first bottleneck is the reliable identification of semiconductors with plausible gap and band-edge energetics. Two-dimensional (2D) materials are relevant to this problem because reduced dimensionality exposes large interfacial area, shortens carrier-transport distances, and allows electronic structure to be tuned through composition, bonding, and dielectric environment.\cite{jin2022jpcl,gao2024acsnano} Databases such as the Computational 2D Materials Database (C2DB) enable such screening across thousands of monolayers generated with a common first-principles workflow.\cite{haastrup2018,gjerding2021}

The database scale required for screening depends on semilocal density-functional theory, and this creates a specific failure mode. PBE systematically underestimates semiconductor band gaps.\cite{pbe} For ordinary ranking this gives a quantitative error, but near the metal--insulator boundary it can become a classification error: a finite-gap semiconductor may be labeled metallic and removed before any photocatalyst criterion is applied. Hybrid functionals such as HSE06 reduce this band-gap bias for many semiconductors,\cite{heyd2003,krukau2006} but database-wide HSE06 calculations remain too expensive for routine use. A practical screening workflow must therefore do two things. It must recover PBE-spurious semiconductors from the nominal metal pool, and it must correct the PBE gaps of the resulting insulating population toward HSE06 accuracy.

$\Delta$-machine learning addresses the second task by predicting the difference between a low-fidelity baseline and a high-fidelity reference rather than predicting the high-fidelity value independently.\cite{ramakrishnan2015} This strategy has been applied successfully in molecular quantum chemistry and is now entering periodic electronic-structure workflows. Examples include HSE06 eigenvalue reconstruction from PBE inputs for crystalline compounds,\cite{adhikari2023} PBE-to-HSE06 correction for metal--nitrogen-doped graphene catalysts,\cite{karimitari2025} and OrbNet-Crystal learning of $\EgHSE-\EgPBE$ on C2DB with HSE06-level band-gap errors near 0.12~eV.\cite{kang2026orbnet} These results show that high-fidelity electronic gaps can be reconstructed from lower-cost information. They do not, by themselves, solve the upstream population problem: a semiconductor-only regressor cannot recover materials that PBE has already excluded as metals.

This upstream error matters for green-hydrogen photocatalyst discovery because candidate membership, not only regression accuracy, controls the final screen. Recent 2D photocatalyst studies combine high-throughput screening with machine-learned gaps, hybrid-functional or many-body validation, optical descriptors, carrier-dynamics estimates, finite-temperature tests, and electrochemical stability analysis.\cite{singh2015,wang2024jpcl,wang2024,tang2026nanoenergy} In such multi-stage workflows, the decisive question is whether the fidelity correction changes which materials pass the physically imposed thresholds. A small global mean absolute error is insufficient if the workflow still removes viable materials upstream or makes incorrect decisions near the gap cutoff.

We address this issue with a leakage-aware two-stage workflow. Stage~I treats PBE-labeled metals with known HSE06 gaps as a classification problem and identifies PBE-spurious metals using only structural, chemical, magnetic, and stability descriptors. HSE06 and GW quantities are excluded as direct target leakage, and PBE electronic quantities are also withheld because they are target-proximal outputs of the same low-fidelity calculation being audited.\cite{kapoor2023} Stage~II then reintroduces PBE electronic quantities as physically meaningful low-fidelity inputs for HSE06--PBE $\Delta$-learning. Thus, a descriptor that would be an inappropriate shortcut in Stage~I becomes the baseline to be corrected in Stage~II. Machine-learning classification of metals and insulators has been explored in other materials contexts,\cite{georgescu2021} but here the classifier is used specifically to repair the PBE-defined screening population before high-fidelity gap correction.

We ask three questions. First, can PBE-spurious metallicity in C2DB be repaired without using electronic leakage features? Second, how accurately can HSE06 band gaps be reconstructed for the corrected insulating population by $\Delta$-learning? Third, does the high-fidelity correction change which 2D materials survive an electronic screen for green-hydrogen photocatalysis? We answer these questions by combining classification, regression, and a bounded pH~0 water-splitting pre-screen. The resulting shortlist is a prioritized validation set for explicit HSE06 band edges, optical response, catalytic kinetics, and stability, not a final claim of photocatalytic performance.

\section{Computational Methods}

\subsection{Data source, labels, and dataset bookkeeping}
All structures and properties were taken from C2DB.\cite{haastrup2018,gjerding2021} No first-principles calculations from the database were recomputed in the present workflow. Of the 16,905 entries in the working C2DB snapshot, 8,686 have a reported PBE gap and form the modeling pool. Using $\EgPBE<0.05$~eV as the operational PBE-metal criterion, this pool contains 4,351 PBE-labeled metals and 4,335 PBE-labeled insulators.

The PBE-metal population is the focus of Stage~I. Among the 4,351 PBE-labeled metals, 159 have an explicit HSE06 gap: 29 remain metallic at HSE06 ($\EgHSE<0.05$~eV), whereas 130 become insulating ($\EgHSE\ge 0.05$~eV). These 159 materials define the labeled classification set, and the remaining 4,192 PBE metals constitute the unlabeled classification target.

A separate bookkeeping point is important for the downstream screen. Across the complete 8,686-material modeling pool, 3,363 entries have an explicit HSE06 gap, comprising 3,318 HSE06 insulators and 45 HSE06 metals. Because only 29 of those 45 HSE06 metals occur in the PBE-metal labeled subset, 16 explicit HSE06 metals occur among PBE-labeled insulators. Accordingly, whenever an HSE06 label is available it takes precedence over the PBE label; the PBE-insulator subset is not treated as infallible. The 3,318 HSE06 insulators form the ground-truth population for Stage~II regression.

\subsection{Stage-specific descriptor construction and leakage control}
We compared three descriptor representations of increasing chemical and structural richness (Table~\ref{tab:featuresets}). The first uses C2DB-native metadata. The second contains Magpie composition statistics generated with \texttt{matminer}, together with basic lattice and density descriptors.\cite{ward2016,ward2018} The third combines these branches with ionic, oxidation-state, and valence-orbital statistics and local coordination descriptors extracted from relaxed structures using \texttt{pymatgen} and the Atomic Simulation Environment.\cite{ong2013,larsen2017}

The raw curated representation contained 841 columns. We used pairwise Pearson correlations only as a redundancy screen for this high-dimensional representation: for pairs with $|r|>0.95$, we removed one member, eliminating 138 columns and leaving 703 columns before stage-specific exclusions. The smaller C2DB-native and Magpie representations were not correlation-pruned. The correlation structure of the compact C2DB-native classification representation is shown in Figure~S1 of the Supporting Information.

We made feature inclusion explicitly stage dependent. In Stage~I, HSE06- and GW-derived quantities are direct target leakage and were removed. We also withheld the PBE gap, direct-gap flag, VBM, CBM, Fermi level, plasma frequency, and half-metal gap. These PBE quantities do not constitute direct HSE06 target leakage, but they are target-proximal electronic shortcuts from the same low-fidelity calculation whose metallic label is being audited. Excluding them makes Stage~I a chemistry/structure-based repair problem rather than a continuation of the PBE electronic label. In Stage~II, by contrast, the PBE gap, vacuum level, Fermi level, VBM, and CBM are physically legitimate low-fidelity inputs because the target is explicitly the HSE06 correction to PBE; only HSE06- and GW-level target quantities were excluded. This stage-specific feature policy produces 9/19 classification/regression features for the C2DB-native representation, 141/141 for Magpie+structural, and 694/702 for the curated representation.

\begin{table}[!ht]
\small
\centering
\caption{Descriptor representations used in the two learning stages. Stage~I excludes HSE06/GW target information to prevent direct leakage and additionally withholds PBE electronic quantities to avoid a low-fidelity shortcut; Stage~II reintroduces PBE electronic quantities as physically legitimate low-fidelity inputs for $\Delta$-learning.}
\label{tab:featuresets}
\begin{tabular}{p{2.45cm}p{7.0cm}cc}
\toprule
Representation & Contents & Class. & Regr. \\
\midrule
C2DB native & Structural, symmetry, stability, magnetic, and thermodynamic metadata native to C2DB; PBE electronic quantities added only for regression & 9 & 19 \\
Magpie + structural & Composition-derived elemental statistics (mean, range, average deviation, and mode of atomic number, electronegativity, valence counts, etc.) plus lattice/density descriptors & 141 & 141 \\
Curated combined & C2DB + Magpie branches, ionic/oxidation-state/valence-orbital statistics, and POSCAR-derived coordination/bond-length descriptors; high-dimensional representation pruned at $|r|>0.95$ & 694 & 702 \\
\bottomrule
\end{tabular}
\end{table}

\subsection{Stage I: PBE-spurious-metal versus true-metal classification}
For the 159 HSE06-labeled PBE metals, the binary target was $y=1$ for a PBE-spurious metal ($\EgHSE\ge0.05$~eV) and $y=0$ for a true HSE06 metal ($\EgHSE<0.05$~eV), matching the final notebook convention used to generate the confusion matrices. We used a stratified 75/25 split (119 training, 40 test; random state 42) and compared four classifiers for each descriptor representation: an RBF support-vector classifier in a standardized pipeline ($C=1.0$); a Random Forest with 50--100 trees, maximum depth 3--5, and \feat{min\_samples\_leaf=5}; Gradient Boosting with 50 estimators, maximum depth 3, learning rate 0.01--0.03, and \feat{min\_samples\_leaf=3}; and XGBoost with 50 estimators, maximum depth 3, and learning rate 0.01--0.03.\cite{xgboost} We median-imputed missing values using statistics fitted only on the training data.

We additionally assessed model stability by 5-fold stratified cross-validation on the training set; the complete comparison is reported in Table~S1. We retained Random Forest as the common classifier because it combines stable cross-validation behavior with the strongest or near-strongest holdout accuracy across the descriptor sets, and because using one classifier family keeps the representation comparison interpretable. Full holdout confusion matrices for all four algorithms are provided in Figure~S2.

We applied the selected classifier to the 4,192 HSE06-unknown PBE metals. Three operational probability bands were used for triage using the predicted PBE-spurious-metal probability: $P_{\mathrm{spur}}\ge0.65$ (predicted PBE-spurious metal, retained as an insulator candidate), $0.35<P_{\mathrm{spur}}<0.65$ (borderline, withheld from the screen), and $P_{\mathrm{spur}}\le0.35$ (predicted true metal). These thresholds are workflow cutoffs rather than calibrated posterior-probability confidence intervals and are interpreted accordingly.

\subsection{Stage II: HSE06--PBE $\Delta$-learning regression}
The regression target was
\begin{equation}
\dEg=\EgHSE-\EgPBE,
\end{equation}
and the reconstructed hybrid-functional gap was
\begin{equation}
\widehat{\EgHSE}=\EgPBE+\widehat{\dEg}.
\end{equation}
We split the 3,318 HSE06-insulator references 80/20 (2,654/664; \feat{random\_state=42}) and screened linear regression, Ridge regression ($\alpha=1$), RBF-SVR, Random Forest, Gradient Boosting, and XGBoost. We subsequently refined XGBoost by manual hyperparameter exploration for each representation. The final regularized models used 200 estimators/depth 3/learning rate 0.05 for C2DB-native descriptors, 1000 estimators/depth 3/learning rate 0.03 for Magpie+structural descriptors, and 1000 estimators/depth 5/learning rate 0.03 for the curated representation; all used \feat{subsample=0.8}, \feat{colsample\_bytree=0.8}, and L1/L2 regularization. We implemented the models with \texttt{scikit-learn} and XGBoost.\cite{pedregosa2011,xgboost}

We retain the fixed 80/20 split to provide a common descriptor-by-descriptor benchmark. The available model-selection workflow does not constitute a fully nested external validation, so the 664-material metrics are best interpreted as internal interpolation benchmarks rather than deployment-level error estimates. For future deployment to chemically distinct material families, repeated/nested model selection and composition- or prototype-grouped validation would provide a more stringent estimate of extrapolative uncertainty.

\subsection{Targeted first-principles validation of ML-predicted gaps}
To complement the database-based test set with material-resolved validation, we performed explicit PBE and HSE06 calculations for six compounds for which ML gap predictions were available and the first-principles calculations were complete: \feat{1AgBr-1}, \feat{1AgCl-1}, \feat{1AgI-1}, \feat{1BrCu-1}, \feat{1ClCu-1}, and \feat{1CuI-1}. The calculations were carried out with VASP~6.4.3 using the projector-augmented-wave method.\cite{kresse1996prb,kresse1996cms,bloechl1994,kresse1999} A plane-wave kinetic-energy cutoff of 550~eV was used, and the Brillouin zone was sampled with Monkhorst--Pack meshes generated from a reciprocal-space spacing of 0.003~\AA$^{-1}$. Periodic images of the monolayers were separated by 15~\AA{} of vacuum. PBE structural relaxations used an electronic self-consistency threshold of $10^{-6}$~eV and were continued until all residual forces were smaller than 0.01~eV~\AA$^{-1}$. The relaxed geometries were then used for spin--orbit-coupled electronic-structure calculations. For each material, a self-consistent PBE+SOC calculation was followed by a single-point HSE06+SOC calculation using the standard HSE06 screened hybrid functional. The independently recalculated PBE gaps agree closely with the corresponding C2DB PBE values (mean absolute difference 0.023~eV; maximum difference 0.05~eV), providing a consistency check before comparison with the new HSE06 references. The explicit HSE06 values are used below as a prospective material-level validation set and, when a screened candidate is affected, to update its screening status.

\subsection{Stage III: electronic pre-screen for overall water splitting}
We merged the Stage~I and Stage~II outputs after giving every explicit HSE06 label precedence over its PBE label. The resulting partition contains 8,221 materials in the HSE06-insulator screening pool (3,318 with DFT-calculated HSE06 gaps and 4,903 with $\Delta$-learned gaps), 57 excluded metals (45 HSE06-confirmed plus 12 predicted metals), and 408 borderline cases. These corrected counts sum to the full 8,686-material modeling pool (Table~\ref{tab:dataset}).

We subjected the insulator pool to three electronic/thermodynamic filters: (i) a practical visible-light gap window of $1.6\le \EgHSE\le2.8$~eV; (ii) a near-hull filter, using $E_{\mathrm{hull}}\le0.05$~eV/atom as the strict criterion and $E_{\mathrm{hull}}\le0.10$~eV/atom as a relaxed metastability tolerance; and (iii) band-edge alignment with the pH~0 water redox potentials on an absolute vacuum scale, requiring $E_{\mathrm{CBM}}\ge-4.44$~eV for $\mathrm{H^+/H_2}$ reduction and $E_{\mathrm{VBM}}\le-5.67$~eV for $\mathrm{O_2/H_2O}$ oxidation.\cite{butler1978,xu2000} The 1.6--2.8~eV interval is a screening choice intended to prioritize visible-light absorbers and should not be confused with the 1.23~eV thermodynamic minimum for overall water splitting.

Because the present model learns only the gap correction and not separate VBM and CBM corrections, we constructed the screened band edges with a symmetric scissor approximation,
\begin{equation}
E_{\mathrm{VBM/CBM}}^{*}=E_{\mathrm{Fermi}}^{\mathrm{abs}}\mp\frac{\widehat{\EgHSE}}{2}.
\end{equation}
This approximation treats the vacuum-referenced Fermi level as a proxy for the gap center and opens the corrected gap symmetrically; it therefore does not capture asymmetric HSE06 shifts of the individual band edges or uncertainty in the assumed center. The resulting shortlist is therefore an electronic-structure pre-screen. Direct HSE06 band edges (or edge-specific $\Delta$-learning), pH dependence, kinetic overpotentials, and photo-/electrochemical stability remain necessary validation steps.

\begin{table}[!ht]
\small
\centering
\caption{Dataset partition used by the corrected workflow. Explicit HSE06 labels take precedence over PBE labels.}
\label{tab:dataset}
\begin{tabular}{lr}
\toprule
Quantity & Count \\
\midrule
Total C2DB entries in working snapshot & 16,905 \\
Entries with PBE gap (modeling pool) & 8,686 \\
\quad PBE insulators ($\EgPBE\ge0.05$~eV) & 4,335 \\
\quad PBE metals ($\EgPBE<0.05$~eV) & 4,351 \\
\qquad HSE06-labeled PBE metals: true / spurious & 159: 29 / 130 \\
\qquad Classification train / test & 119 / 40 \\
\qquad HSE06-unknown PBE metals classified downstream & 4,192 \\
Materials with explicit HSE06 gap (any PBE label) & 3,363 \\
\quad HSE06 insulators / metals & 3,318 / 45 \\
\quad $\Delta$-learning train / test & 2,654 / 664 \\
Final HSE06-insulator screening pool & 8,221 \\
\quad DFT-HSE06 / ML-HSE06 gaps & 3,318 / 4,903 \\
Excluded HSE06-confirmed + predicted metals & 57 \\
Borderline classifier cases (withheld) & 408 \\
\bottomrule
\end{tabular}
\end{table}

\begin{figure}[!ht]
\centering
\begin{tikzpicture}[
  node distance=7mm and 10mm,
  box/.style={draw, rounded corners, align=center, minimum height=8mm, minimum width=33mm, font=\small, fill=blue!5},
  fset/.style={draw, rounded corners, align=center, minimum height=8mm, minimum width=30mm, font=\small, fill=orange!10},
  decision/.style={draw, rounded corners, align=center, minimum height=8mm, font=\small, fill=green!8},
  arr/.style={-{Latex[length=2mm]}, thick}
]
\node[box] (c2db) {C2DB working snapshot\\ $N=16{,}905$};
\node[box, below=of c2db] (pbe) {PBE gap available\\ $N=8{,}686$};
\node[box, below left=9mm and -4mm of pbe] (pbeins) {PBE insulators\\ $N=4{,}335$};
\node[box, below right=9mm and -4mm of pbe] (pbemet) {PBE metals\\ $N=4{,}351$};
\node[box, below=of pbemet] (known) {HSE06-labeled subset\\ $N=159$ (29/130)};
\node[fset, below=16mm of known] (f2) {Magpie + structural};
\node[fset, left=15mm of f2] (f1) {C2DB native};
\node[fset, right=15mm of f2] (f3) {Curated combined};
\node[decision, below=14mm of f2, minimum width=61mm] (clf) {4 classifiers; leakage-controlled features\\CV and holdout evaluation};
\node[box, below=9mm of clf] (unlabeled) {HSE06-unknown PBE metals\\ $N=4{,}192$};
\node[box, below=8mm of unlabeled, minimum width=66mm] (tiers) {Operational probability triage: insulator / borderline / metal};
\draw[arr] (c2db) -- (pbe);
\draw[arr] (pbe) -| (pbeins);
\draw[arr] (pbe) -| (pbemet);
\draw[arr] (pbemet) -- (known);
\draw[arr] (known) -- (f1);
\draw[arr] (known) -- (f2);
\draw[arr] (known) -- (f3);
\draw[arr] (f1) -- (clf.north west);
\draw[arr] (f2) -- (clf.north);
\draw[arr] (f3) -- (clf.north east);
\draw[arr] (clf) -- (unlabeled);
\draw[arr] (unlabeled) -- (tiers);
\end{tikzpicture}
\caption{Stage~I workflow for triaging PBE-predicted metals. HSE06/GW target information is excluded to prevent direct leakage, while PBE electronic quantities are withheld so that classification tests whether non-electronic materials information can repair the low-fidelity metallic label. Explicit HSE06 labels anywhere in the 8,686-material pool supersede PBE labels downstream.}
\label{fig:pipeline1}
\end{figure}
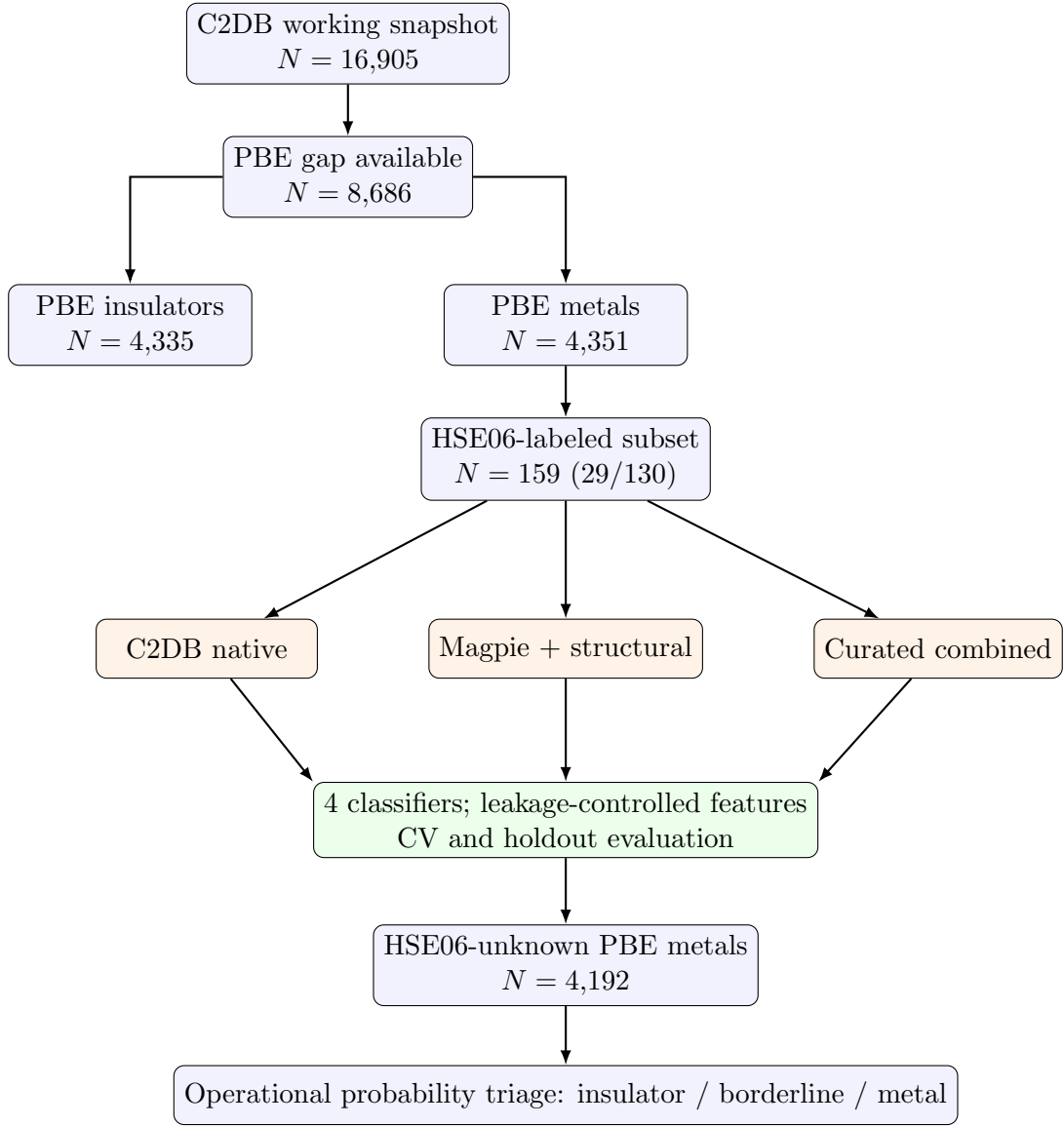

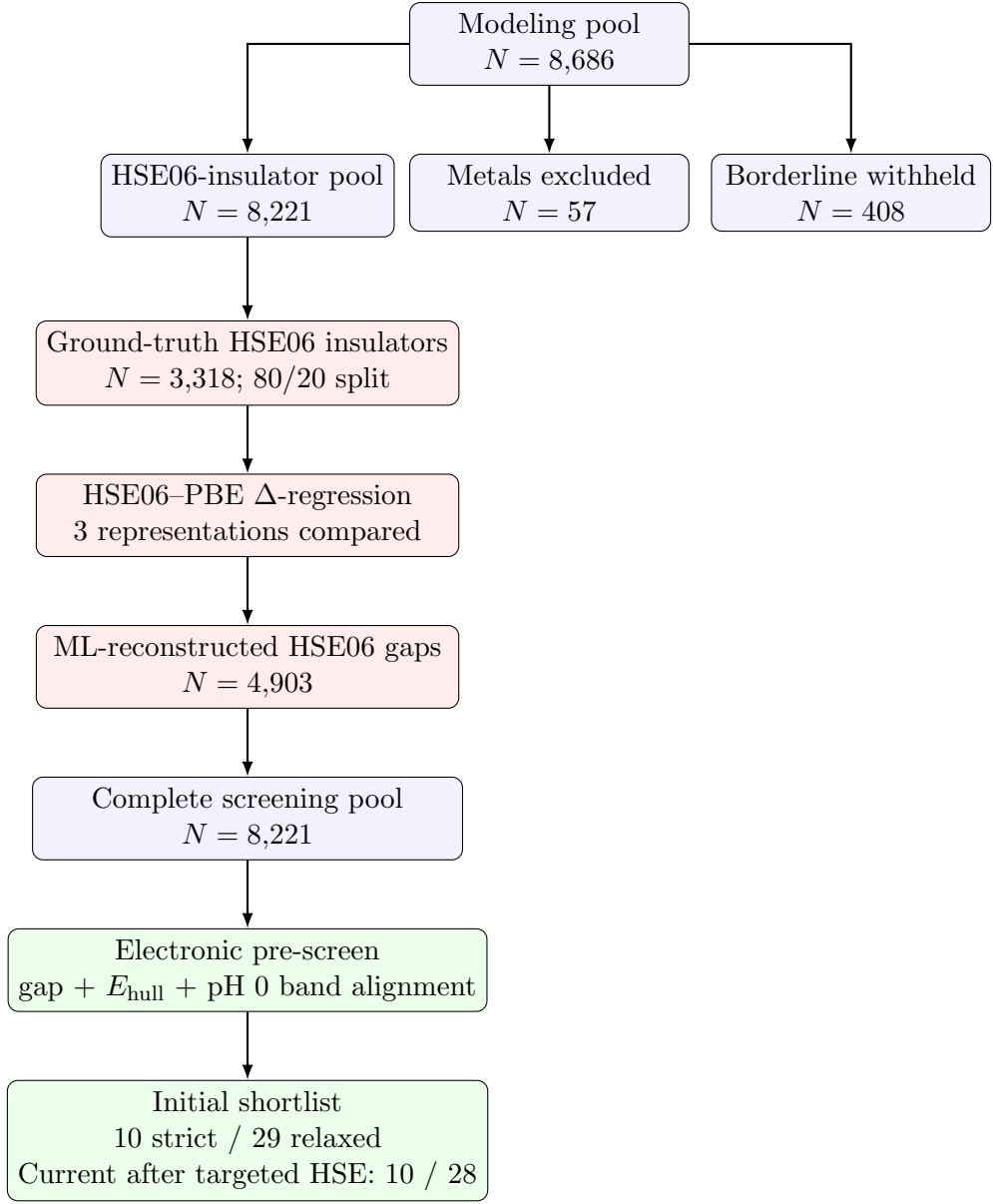
\begin{figure}[!ht]
\centering
\begin{tikzpicture}[
  node distance=7mm and 10mm,
  box/.style={draw, rounded corners, align=center, minimum height=8mm, minimum width=37mm, font=\small, fill=blue!5},
  reg/.style={draw, rounded corners, align=center, minimum height=8mm, minimum width=56mm, font=\small, fill=red!7},
  filt/.style={draw, rounded corners, align=center, minimum height=8mm, minimum width=58mm, font=\small, fill=green!8},
  arr/.style={-{Latex[length=2mm]}, thick}
]
\node[box] (pool) {Modeling pool\\ $N=8{,}686$};
\node[box, below=9mm of pool, xshift=-40mm] (ins) {HSE06-insulator pool\\ $N=8{,}221$};
\node[box, below=9mm of pool] (met) {Metals excluded\\ $N=57$};
\node[box, below=9mm of pool, xshift=40mm] (bord) {Borderline withheld\\ $N=408$};
\node[reg, below=11mm of ins] (gt) {Ground-truth HSE06 insulators\\$N=3{,}318$; 80/20 split};
\node[reg, below=9mm of gt] (xgb) {HSE06--PBE $\Delta$-regression\\3 representations compared};
\node[reg, below=9mm of xgb] (pred) {ML-reconstructed HSE06 gaps\\$N=4{,}903$};
\node[box, below=9mm of pred, minimum width=57mm] (merged) {Complete screening pool\\ $N=8{,}221$};
\node[filt, below=9mm of merged] (filter) {Electronic pre-screen\\gap + $E_{\rm hull}$ + pH~0 band alignment};
\node[filt, below=9mm of filter] (cand) {Initial shortlist\\10 strict / 29 relaxed\\Current after targeted HSE: 10 / 28};
\draw[arr] (pool) -| (ins);
\draw[arr] (pool) -- (met);
\draw[arr] (pool) -| (bord);
\draw[arr] (ins) -- (gt);
\draw[arr] (gt) -- (xgb);
\draw[arr] (xgb) -- (pred);
\draw[arr] (pred) -- (merged);
\draw[arr] (merged) -- (filter);
\draw[arr] (filter) -- (cand);
\end{tikzpicture}
\caption{Stages~II--III: $\Delta$-learning of the HSE06--PBE band-gap correction followed by the water-splitting electronic pre-screen. The initial screen yields 10 strict and 29 relaxed candidates. Subsequent targeted HSE06 validation removes one relaxed candidate (\feat{1AgBr-1}), giving a current retained set of 10 strict and 28 relaxed materials.}
\label{fig:pipeline2}
\end{figure}

\FloatBarrier

\section{Results and Discussion}

\subsection{Classification remains minority-class limited despite descriptor enrichment}
Table~\ref{tab:classification} summarizes the selected Random Forest classifier on the common 40-material holdout set. The raw accuracy varies only from 85.0 to 87.5\%. This number should not be interpreted in isolation: the holdout set contains 33 PBE-spurious metals and only 7 true HSE06 metals, so an always-insulator classifier already attains 82.5\% accuracy. Balanced accuracy is therefore more informative, increasing from 0.571 for the C2DB-native representation to 0.643 for both Magpie+structural and curated combined. The C2DB-native model recovers 1 of the 7 true metals, while the enriched representations recover 2 of the 7; all three avoid assigning any PBE-spurious metal to the true-metal class. The selected models are consequently useful as a triage device but not as a definitive metal/insulator classifier.

\begin{table}[!ht]
\small
\centering
\caption{Stage~I performance of the selected Random Forest classifiers on the 40-material holdout set. Precision, recall, and F1 refer to the minority true-HSE06-metal class. The cross-validation values are mean $\pm$ standard deviation on the training folds.}
\label{tab:classification}
\resizebox{\textwidth}{!}{%
\begin{tabular}{lcccccccc}
\toprule
Representation & Test acc. & 5-fold CV acc. & Precision & Recall & F1 & Specificity & Balanced acc. & (TN,FP,FN,TP) \\
\midrule
C2DB native & 85.0\% & $81.5\pm1.9\%$ & 1.000 & 0.143 & 0.250 & 1.000 & 0.571 & (33,0,6,1) \\
Magpie + structural & 87.5\% & $83.2\pm2.5\%$ & 1.000 & 0.286 & 0.444 & 1.000 & 0.643 & (33,0,5,2) \\
Curated combined & 87.5\% & $82.4\pm3.0\%$ & 1.000 & 0.286 & 0.444 & 1.000 & 0.643 & (33,0,5,2) \\
\bottomrule
\end{tabular}%
}
\end{table}

\begin{figure}[!ht]
\centering
\begin{subfigure}{0.32\textwidth}
\includegraphics[width=\linewidth]{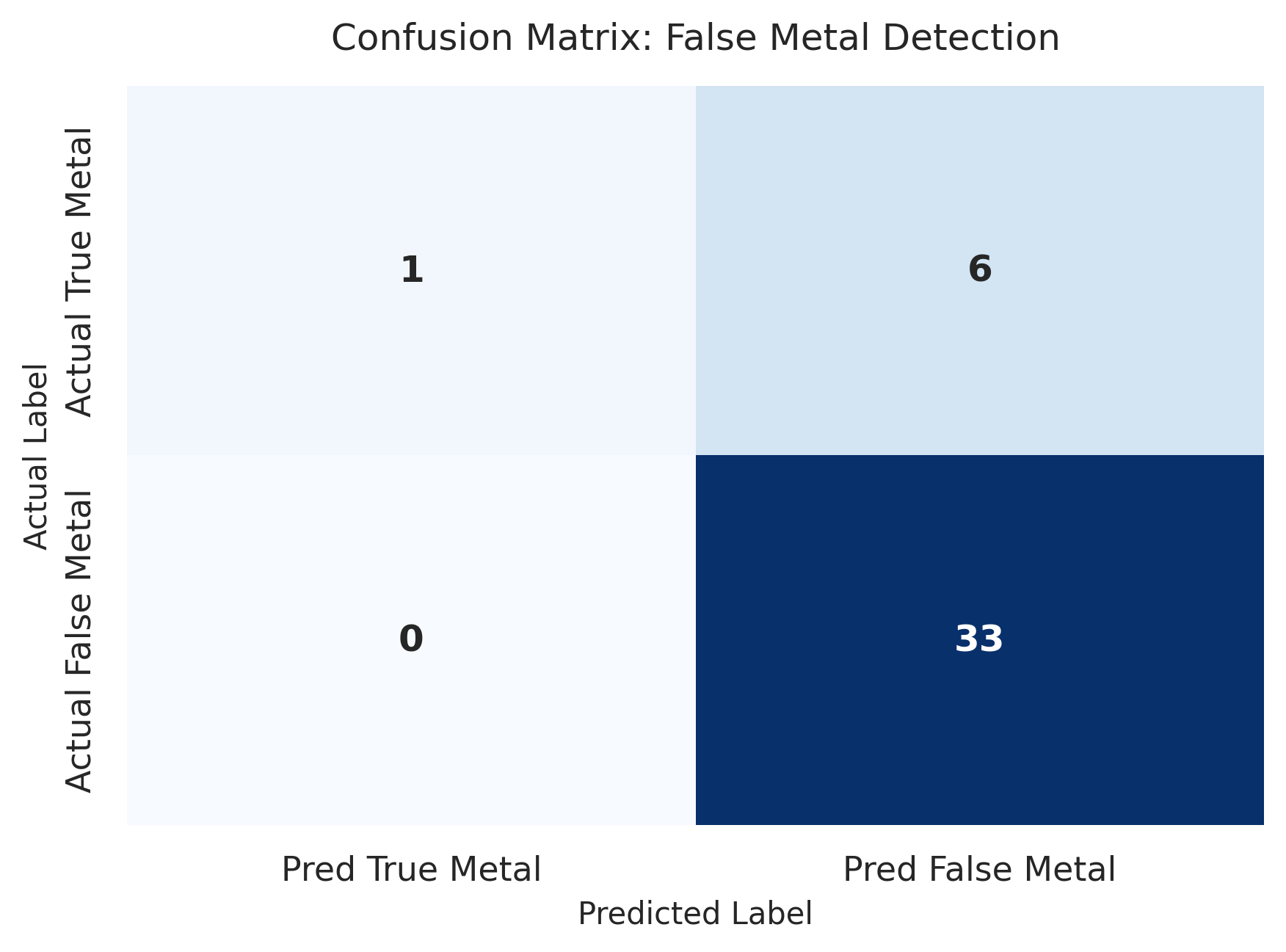}
\caption{C2DB native}
\end{subfigure}
\begin{subfigure}{0.32\textwidth}
\includegraphics[width=\linewidth]{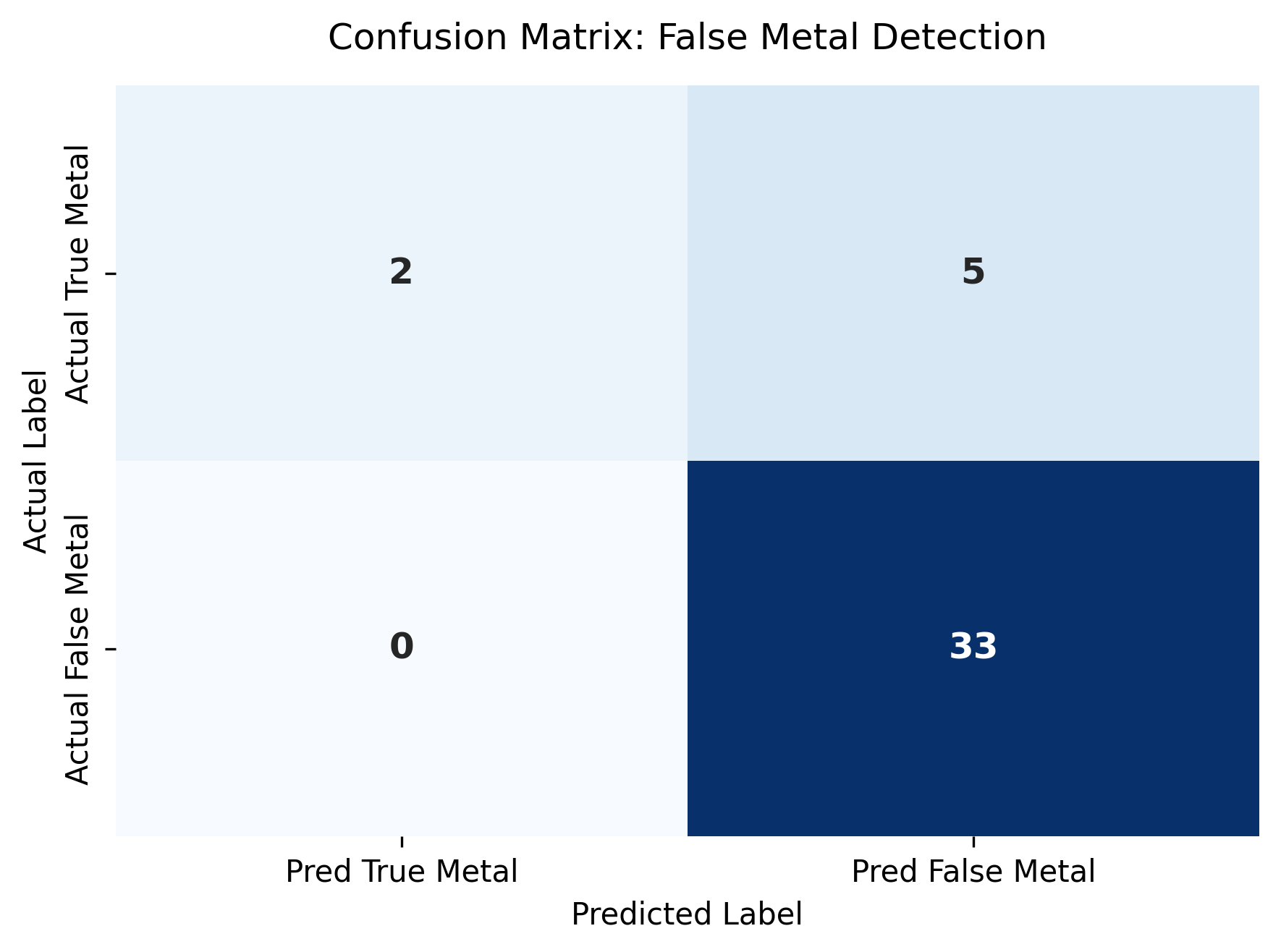}
\caption{Magpie + structural}
\end{subfigure}
\begin{subfigure}{0.32\textwidth}
\includegraphics[width=\linewidth]{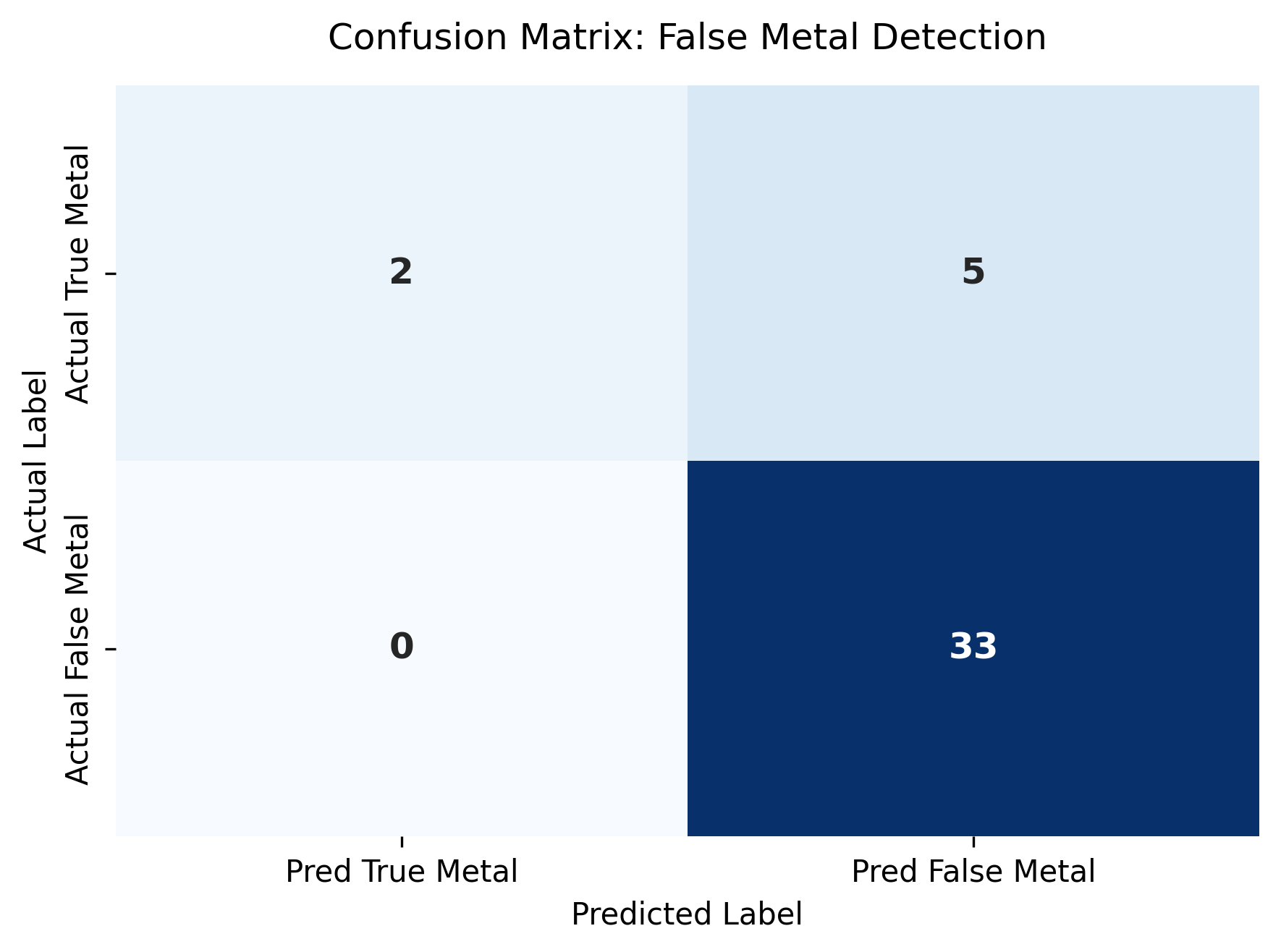}
\caption{Curated combined}
\end{subfigure}
\caption{Holdout confusion matrices for the selected Random Forest classifiers. Although overall accuracy is 85.0--87.5\%, the C2DB-native representation recovers 1 of the 7 true HSE06 metals, while the enriched representations recover 2 of the 7; the figure therefore visualizes why this stage is used for triage rather than definitive classification. The majority class contains 33 PBE-spurious metals (labeled ``Insulator'').}
\label{fig:rfconf}
\end{figure}

Figure~\ref{fig:rfconf} makes the class imbalance explicit: descriptor enrichment increases true-metal recovery from 1 to 2 materials, but the minority class remains poorly sampled. The Supporting Information contains the corresponding training-set matrices, cross-validation comparison, and extended regression diagnostics; together they show that the principal limitation is minority-class statistics rather than descriptor dimensionality.

\begin{figure}[!ht]
\centering
\begin{subfigure}{0.48\textwidth}
\includegraphics[width=\linewidth]{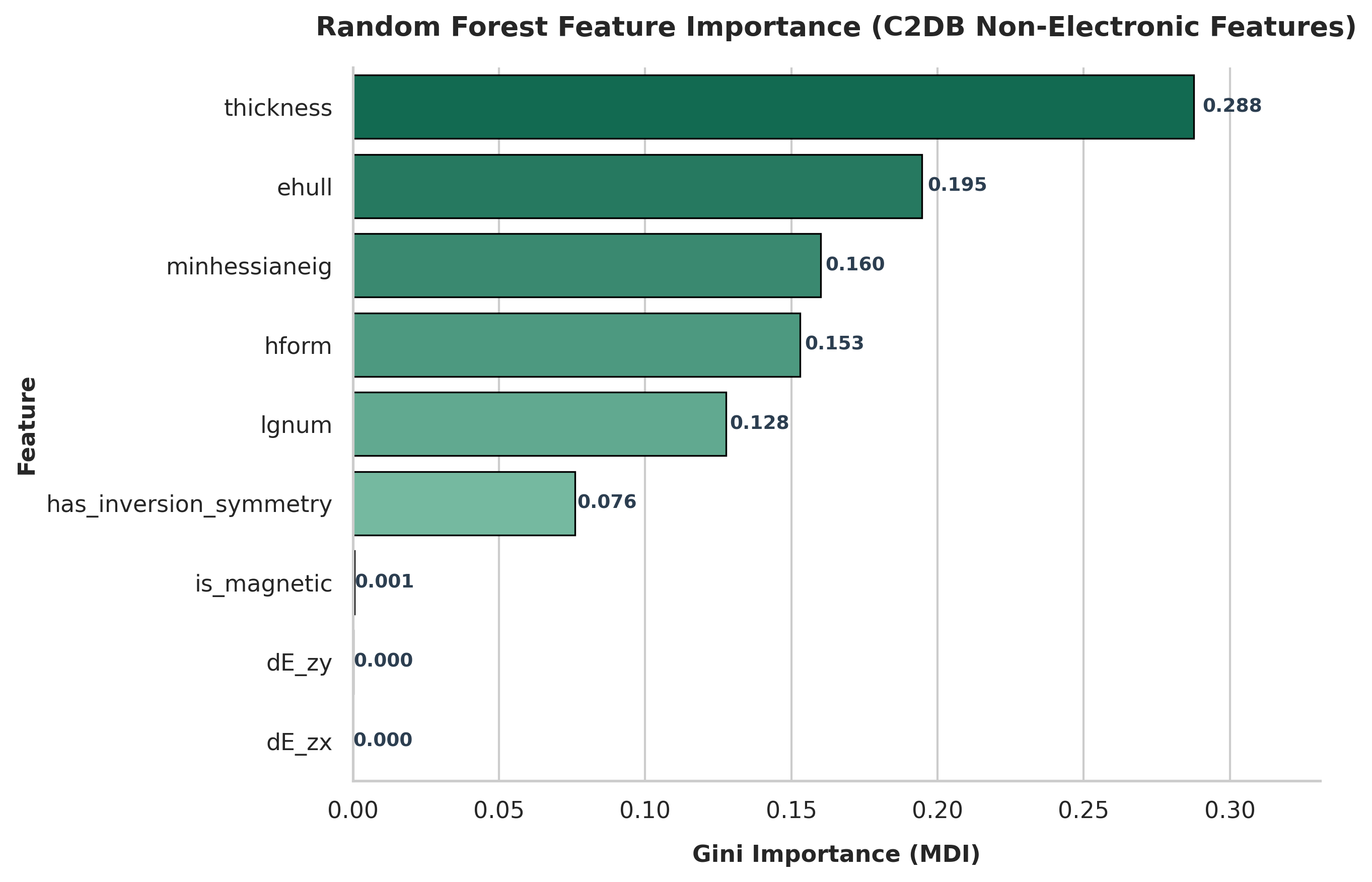}
\caption{C2DB native (9 features)}
\end{subfigure}
\begin{subfigure}{0.48\textwidth}
\includegraphics[width=\linewidth]{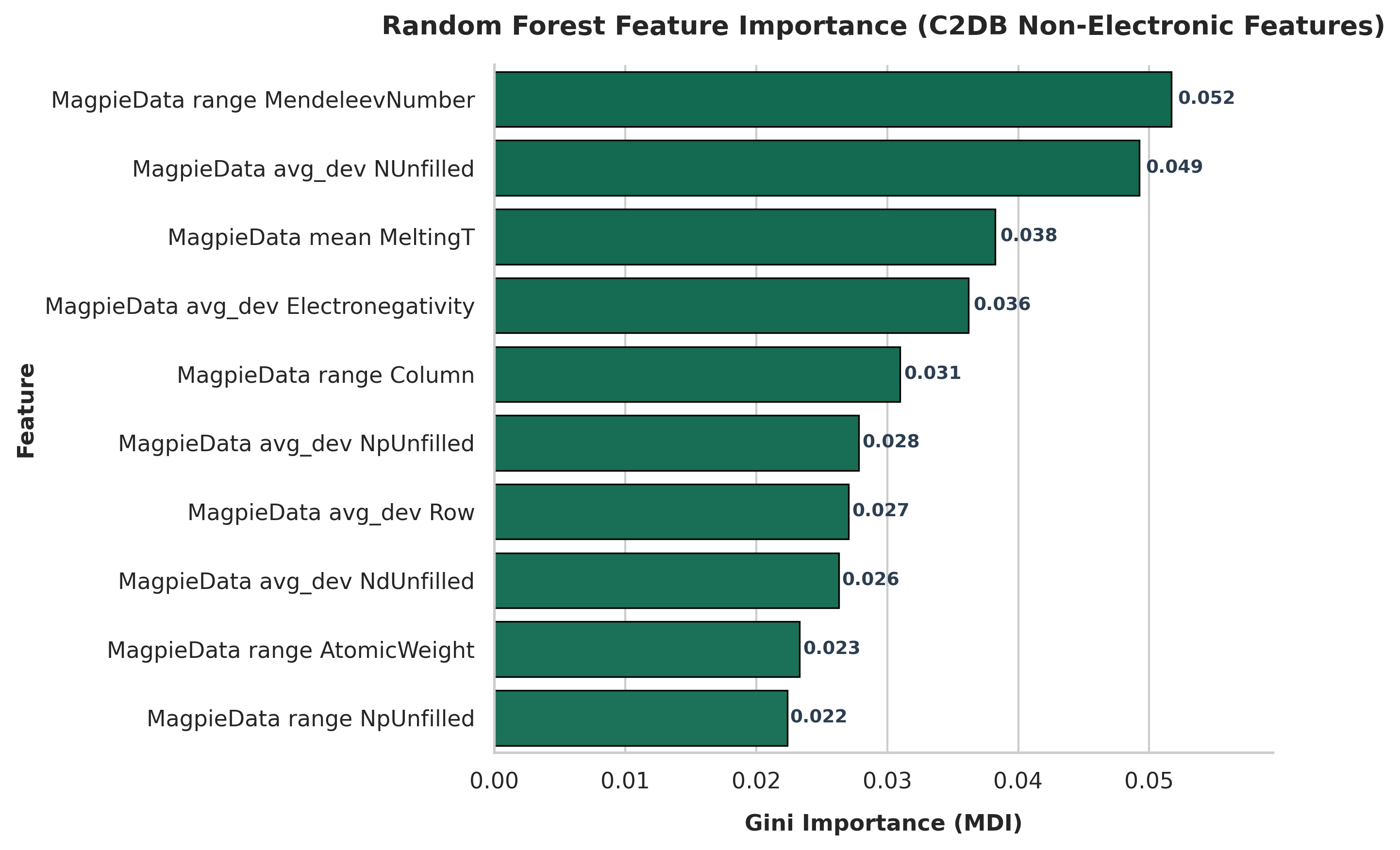}
\caption{Magpie + structural}
\end{subfigure}
\begin{subfigure}{0.62\textwidth}
\includegraphics[width=\linewidth]{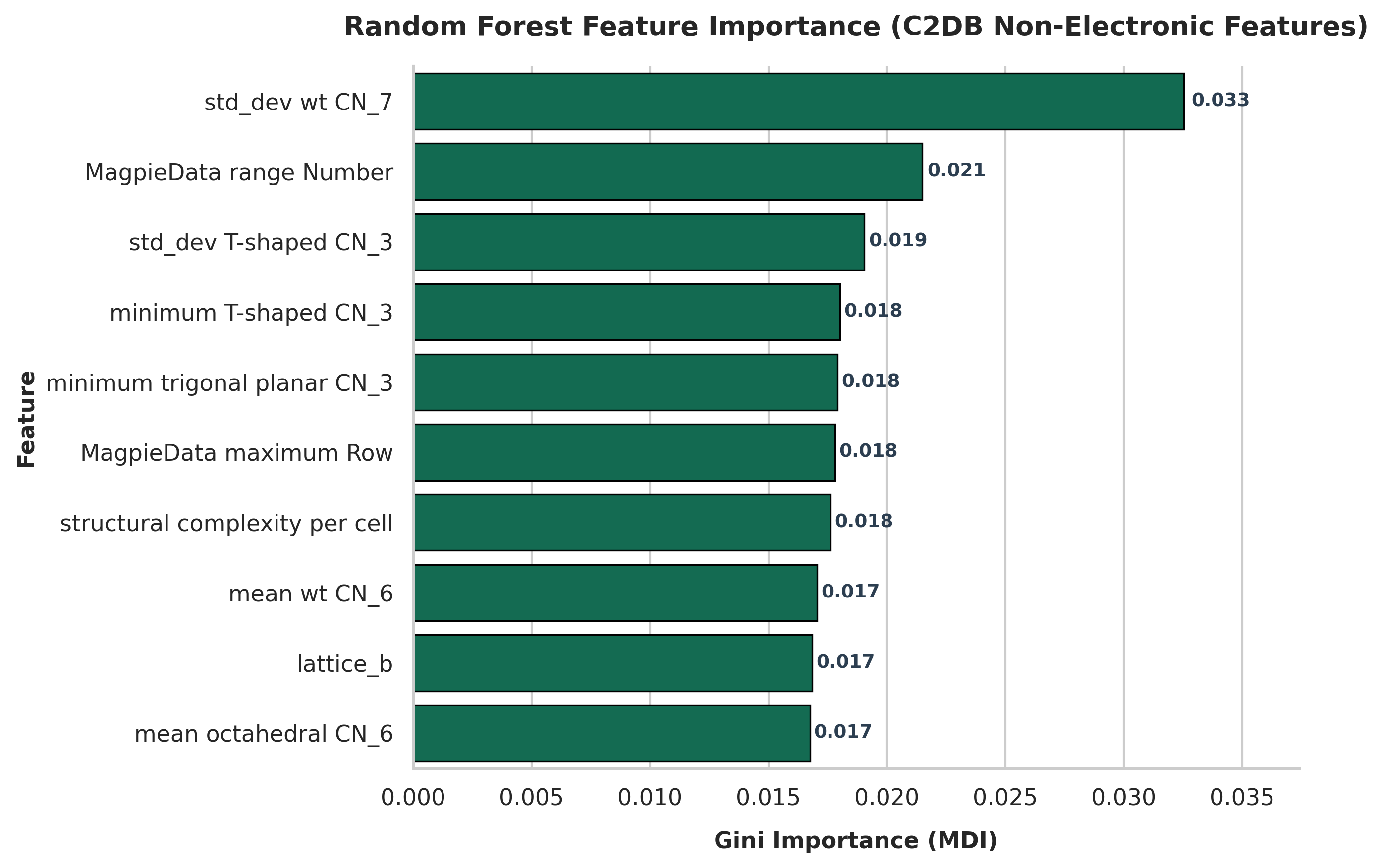}
\caption{Curated combined (top 20 of 694)}
\end{subfigure}
\caption{Random Forest Gini feature importance for true-HSE06-metal versus PBE-spurious-metal classification. The compact representation emphasizes thickness, symmetry, and thermodynamic descriptors, the Magpie representation emphasizes composition-derived statistics, and the curated representation distributes weight across local coordination, bond-length, and lattice descriptors. Importances are interpreted qualitatively because correlated features can redistribute impurity-based weight.}
\label{fig:featimp_clf}
\end{figure}

The feature-importance profiles nevertheless reveal physically interpretable differences (Figure~\ref{fig:featimp_clf}). In the compact C2DB-native model, \feat{thickness} has the largest Gini importance (about 22\%), followed by the layer-group number, formation enthalpy, minimum Hessian eigenvalue, and energy above hull. These variables collectively encode confinement, symmetry, and structural/thermodynamic environment. The Magpie representation highlights composition-derived trends, whereas in the 694-feature curated representation importance is distributed over local coordination motifs, bond-length statistics, and lattice descriptors, with no single feature dominating. Because impurity-based importance can redistribute weight among correlated or high-cardinality descriptors, these rankings are best read as qualitative indicators of the information used by the model, not as causal measures of what determines metallicity.

\FloatBarrier

\subsection{Curated descriptors reduce PBE-to-HSE06 gap error by approximately 90\%}
Stage~II is statistically better supported than Stage~I because 3,318 HSE06-insulator references are available. Table~\ref{tab:regression} compares the six regressors. Since $\widehat{\EgHSE}=\EgPBE+\widehat{\dEg}$, the absolute error in $\widehat{\EgHSE}$ is identical to the error in $\dEg$ for each material; the MAE is therefore the same for both targets, whereas $R^2$ differs because their variances differ.

Unregularized linear regression fails for the 702-feature curated matrix, whereas Ridge regression restores a physically reasonable MAE of 0.198~eV, confirming that the instability is estimator-specific. The nonlinear models benefit consistently from descriptor enrichment: the final regularized XGBoost models give MAEs of 0.239, 0.197, and 0.108~eV for C2DB-native, Magpie+structural, and curated representations, respectively. The curated model reaches $R^2_{\dEg}=0.906$ and $R^2_{\mathrm{abs}}=0.989$, reducing the 1.036~eV raw-PBE error by approximately 90\%. This accuracy is comparable to the emerging periodic-materials $\Delta$-learning literature, including HSE06 eigenvalue reconstruction and the recent OrbNet-Crystal C2DB gap model, although direct numerical ranking is inappropriate because the targets, splits, and representations differ.\cite{adhikari2023,karimitari2025,kang2026orbnet} The important distinction here is that the regressor operates inside a population-repair and photocatalyst-screening workflow rather than as a stand-alone gap predictor.

\begin{table}[!ht]
\small
\centering
\caption{Stage~II $\Delta$-learning regression on the 664-material test set. MAE is identical for $\dEg$ and the reconstructed absolute HSE06 gap by construction; $R^2$ is reported separately because the target variances differ.}
\label{tab:regression}
\begin{tabular}{llccc}
\toprule
Representation & Model & MAE (eV) & $R^2_{\dEg}$ & $R^2_{\rm abs.\ HSE06}$ \\
\midrule
\multirow{6}{*}{C2DB native (19)}
 & Linear Regression & 0.279 & 0.498 & 0.941 \\
 & Ridge Regression & 0.279 & 0.499 & 0.941 \\
 & SVR (RBF) & 0.213 & 0.677 & 0.962 \\
 & Random Forest & 0.224 & 0.670 & 0.961 \\
 & Gradient Boosting & 0.211 & 0.712 & 0.966 \\
 & XGBoost (regularized final) & 0.239 & 0.641 & 0.958 \\
\midrule
\multirow{6}{*}{Magpie + structural (141)}
 & Linear Regression & 0.292 & 0.557 & 0.937 \\
 & Ridge Regression & 0.294 & 0.555 & 0.937 \\
 & SVR (RBF) & 0.192 & 0.755 & 0.965 \\
 & Random Forest & 0.194 & 0.749 & 0.965 \\
 & Gradient Boosting & 0.165 & 0.807 & 0.973 \\
 & XGBoost (regularized final) & 0.197 & 0.770 & 0.968 \\
\midrule
\multirow{6}{*}{Curated combined (702)}
 & Linear Regression & 14.563 & $-2.10\times10^{5}$ & $-2.47\times10^{4}$ \\
 & Ridge Regression & 0.198 & 0.749 & 0.971 \\
 & SVR (RBF) & 0.166 & 0.796 & 0.976 \\
 & Random Forest & 0.143 & 0.846 & 0.982 \\
 & Gradient Boosting & 0.114 & 0.897 & 0.988 \\
 & XGBoost (regularized final) & \textbf{0.108} & \textbf{0.906} & \textbf{0.989} \\
\bottomrule
\end{tabular}
\end{table}

\subsubsection{Material-resolved validation against new HSE06 calculations}
Table~\ref{tab:spotvalidation} compares the new explicit HSE06 gaps with the available model predictions for all six compounds in the completed validation set. The PBE baseline is poor for these systems (MAE 1.480~eV), whereas the C2DB-native, Magpie+structural, and curated models reduce the MAE to 0.417, 0.182, and 0.110~eV, respectively. The curated model is therefore also the most accurate on this small prospective validation set, although the six-material size should be kept in mind when comparing representation-level performance.

\begin{table}[!ht]
\scriptsize
\centering
\caption{Material-resolved validation using explicit HSE06 calculations performed in this work. Values in parentheses are absolute errors relative to the new HSE06 reference. All six materials with completed HSE06 calculations and available ML predictions are retained.}
\label{tab:spotvalidation}
\resizebox{\textwidth}{!}{%
\begin{tabular}{lccccc}
\toprule
Material & HSE06 this work & PBE this work & C2DB-native ML & Magpie ML & Curated ML \\
 & (eV) & (eV; $|\epsilon|$) & (eV; $|\epsilon|$) & (eV; $|\epsilon|$) & (eV; $|\epsilon|$) \\
\midrule
\feat{1AgBr-1} & 2.97 & 1.62 (1.35) & 2.65 (0.32) & 2.70 (0.27) & 2.78 (0.19) \\
\feat{1AgCl-1} & 3.16 & 1.68 (1.48) & 2.75 (0.41) & 2.85 (0.31) & 3.03 (0.13) \\
\feat{1AgI-1}  & 2.70 & 1.55 (1.15) & 2.58 (0.12) & 2.60 (0.10) & 2.55 (0.15) \\
\feat{1BrCu-1} & 2.72 & 1.03 (1.69) & 2.14 (0.58) & 2.83 (0.11) & 2.74 (0.02) \\
\feat{1ClCu-1} & 2.84 & 1.09 (1.75) & 2.21 (0.63) & 3.00 (0.16) & 2.85 (0.01) \\
\feat{1CuI-1}  & 2.75 & 1.29 (1.46) & 2.31 (0.44) & 2.89 (0.14) & 2.91 (0.16) \\
\midrule
MAE & --- & 1.480 & 0.417 & 0.182 & 0.110 \\
\bottomrule
\end{tabular}%
}
\end{table}

The close agreement between the recalculated and C2DB PBE gaps (mean absolute difference 0.023~eV) indicates that the large PBE--HSE06 shifts are not caused by an obvious mismatch in the PBE baseline. The validation set also illustrates why pointwise accuracy and screening decisions are distinct: \feat{1AgBr-1} is predicted within about 0.2~eV by the available Magpie model, yet the explicit HSE06 gap of 2.97~eV crosses the 2.8~eV screening threshold. By contrast, \feat{1AgI-1} remains inside the window at 2.70~eV. This threshold crossing is incorporated into the updated relaxed shortlist discussed below.

\begin{figure}[!ht]
\centering
\includegraphics[width=0.62\linewidth]{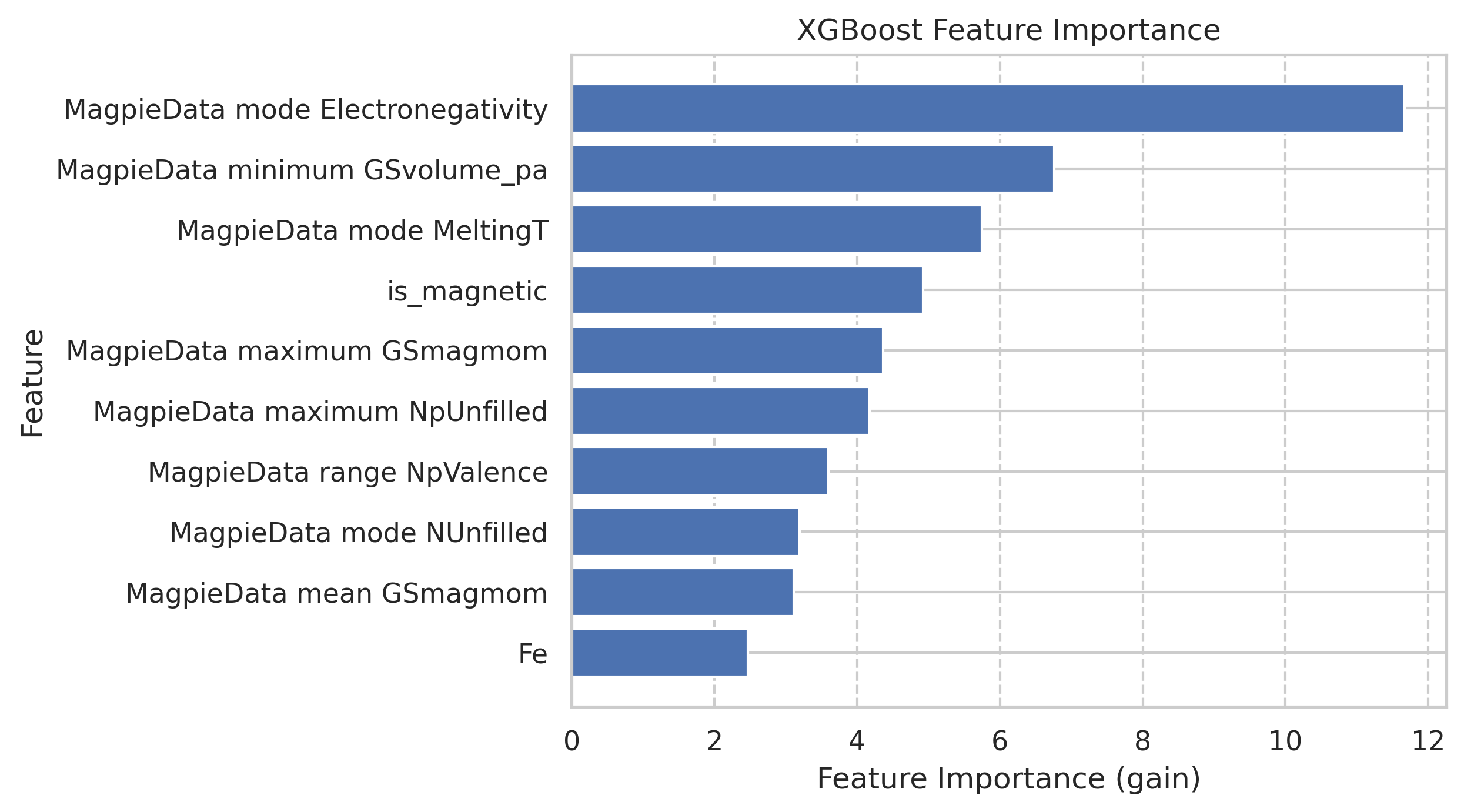}
\caption{Gain-based XGBoost feature importance for the curated $\Delta$-learning regressor (top 10 of 702 features). The leading descriptors span chemical-scale statistics (electronegativity, atomic volume, melting temperature) and native C2DB information, consistent with complementary rather than single-feature control of the HSE06 correction.}
\label{fig:featimp_reg}
\end{figure}

The curated XGBoost importance profile (Figure~\ref{fig:featimp_reg}) is led by Magpie mode electronegativity and minimum ground-state volume per atom, followed by melting-temperature statistics and the C2DB magnetic flag. The corresponding C2DB-native and Magpie-only rankings are provided in Figure~S4. The recurrence of electronegativity and volume descriptors across the Magpie and curated models suggests that chemical scale and bonding environment contain substantial information about the semilocal-to-hybrid correction. The curated model then gains additional predictive capacity from complementary structural and native-C2DB descriptors rather than replacing the composition signal altogether. As with the classifier, gain-based importance is interpreted descriptively rather than causally.

\begin{figure}[!ht]
\centering
\begin{subfigure}{0.48\textwidth}
\includegraphics[width=\linewidth]{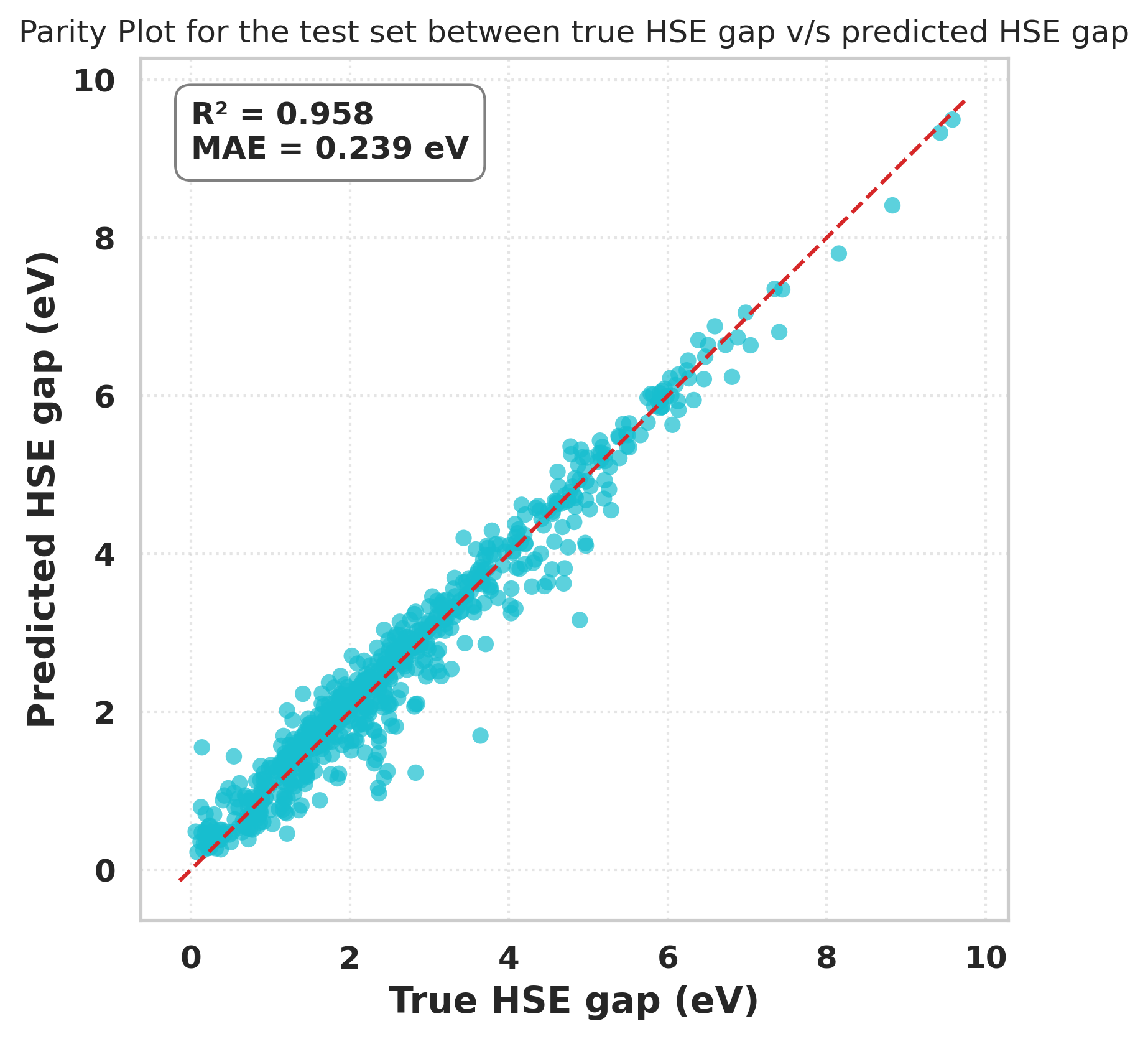}
\caption{C2DB native}
\end{subfigure}
\begin{subfigure}{0.48\textwidth}
\includegraphics[width=\linewidth]{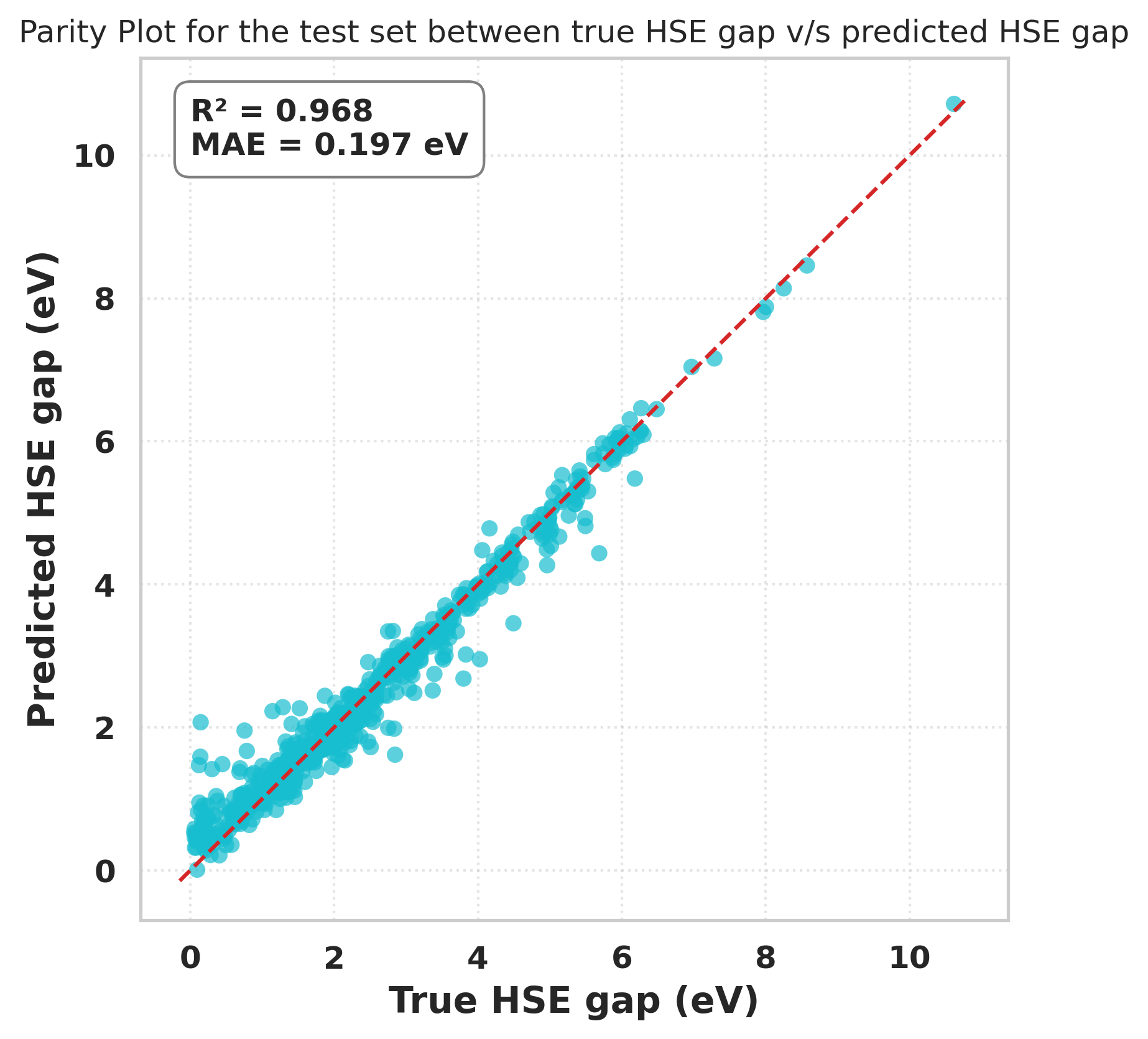}
\caption{Magpie + structural}
\end{subfigure}
\begin{subfigure}{0.48\textwidth}
\includegraphics[width=\linewidth]{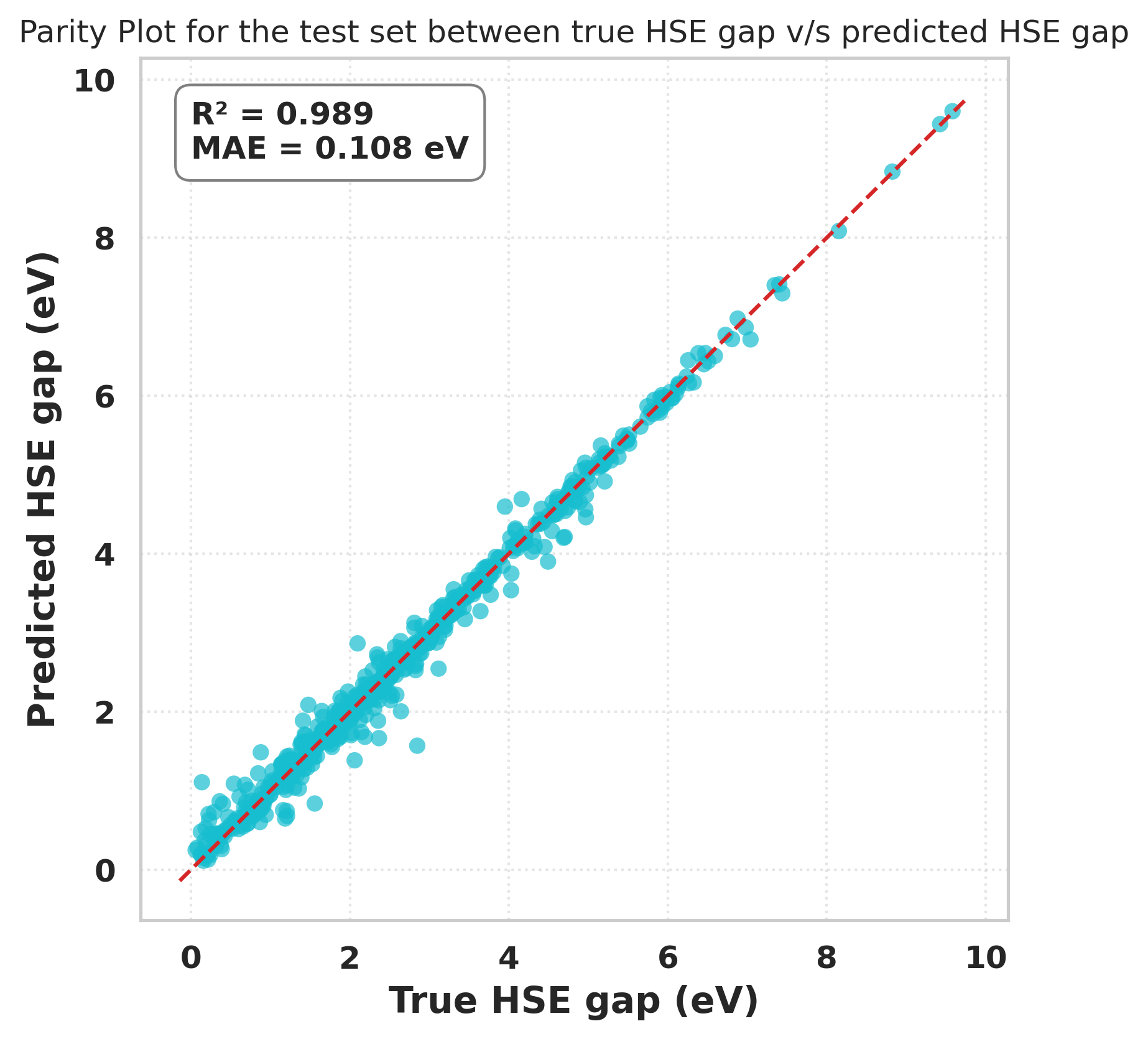}
\caption{Curated combined}
\end{subfigure}
\caption{Regression performance on the 664-material test sets. Panels (a--c) compare reconstructed and reference HSE06 gaps for the final regularized XGBoost models. The curated representation reduces the mean absolute error to 0.108~eV, approximately 90\% below the 1.036~eV raw-PBE baseline.}
\label{fig:parity}
\end{figure}

Figure~\ref{fig:parity} makes the pointwise accuracy explicit. The parity panels show progressively narrower scatter from C2DB native to Magpie and curated descriptors, with the curated model remaining especially tight in the low-to-intermediate gap region most relevant to visible-light screening. The MAE values in Table~\ref{tab:regression} place this improvement against the raw-PBE baseline: the key gain is not a small optimization among ML models, but an order-of-magnitude reduction in the dominant low-fidelity band-gap error.

\begin{figure}[!ht]
\centering
\begin{subfigure}{0.32\textwidth}
\includegraphics[width=\linewidth]{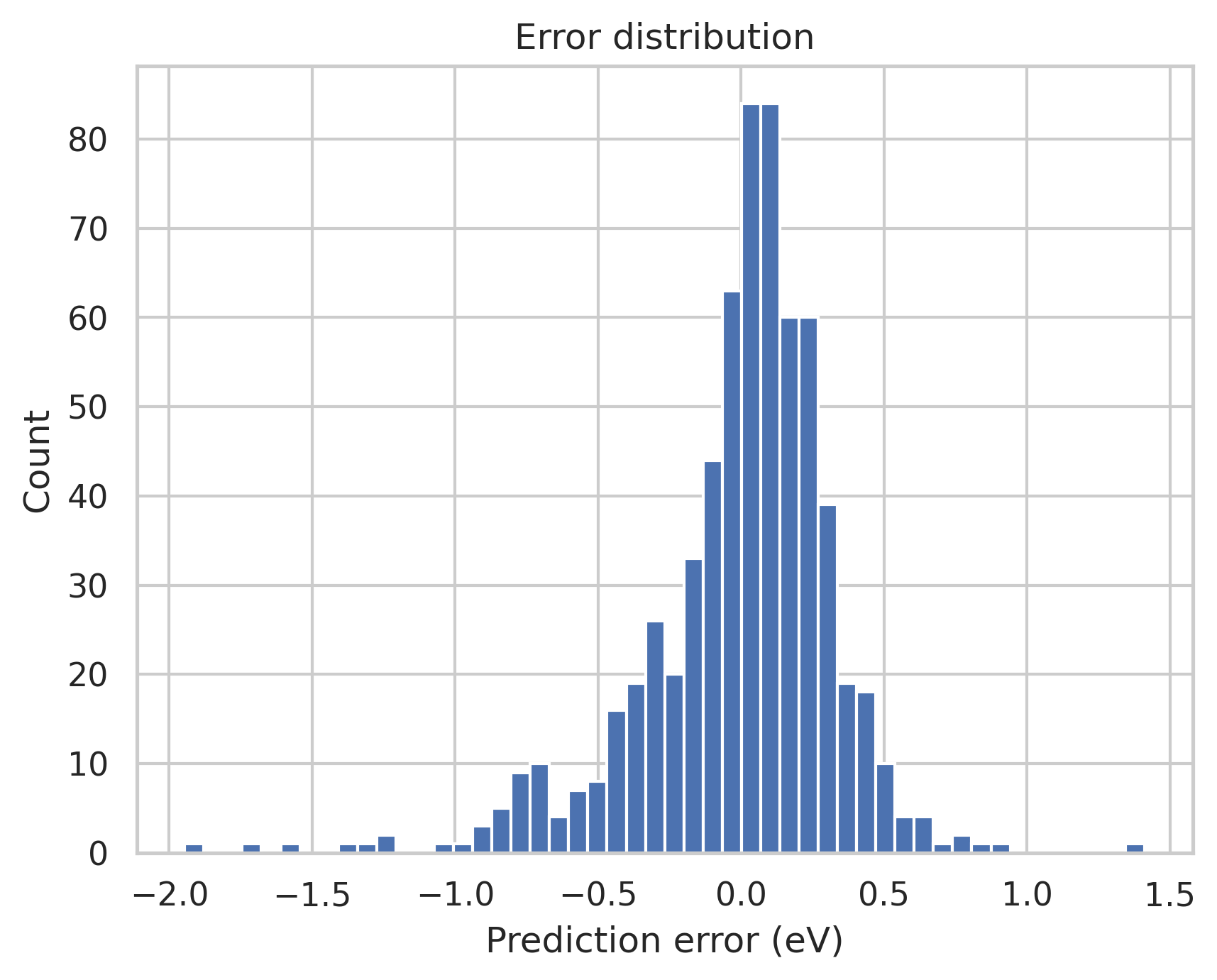}
\caption{C2DB native}
\end{subfigure}
\begin{subfigure}{0.32\textwidth}
\includegraphics[width=\linewidth]{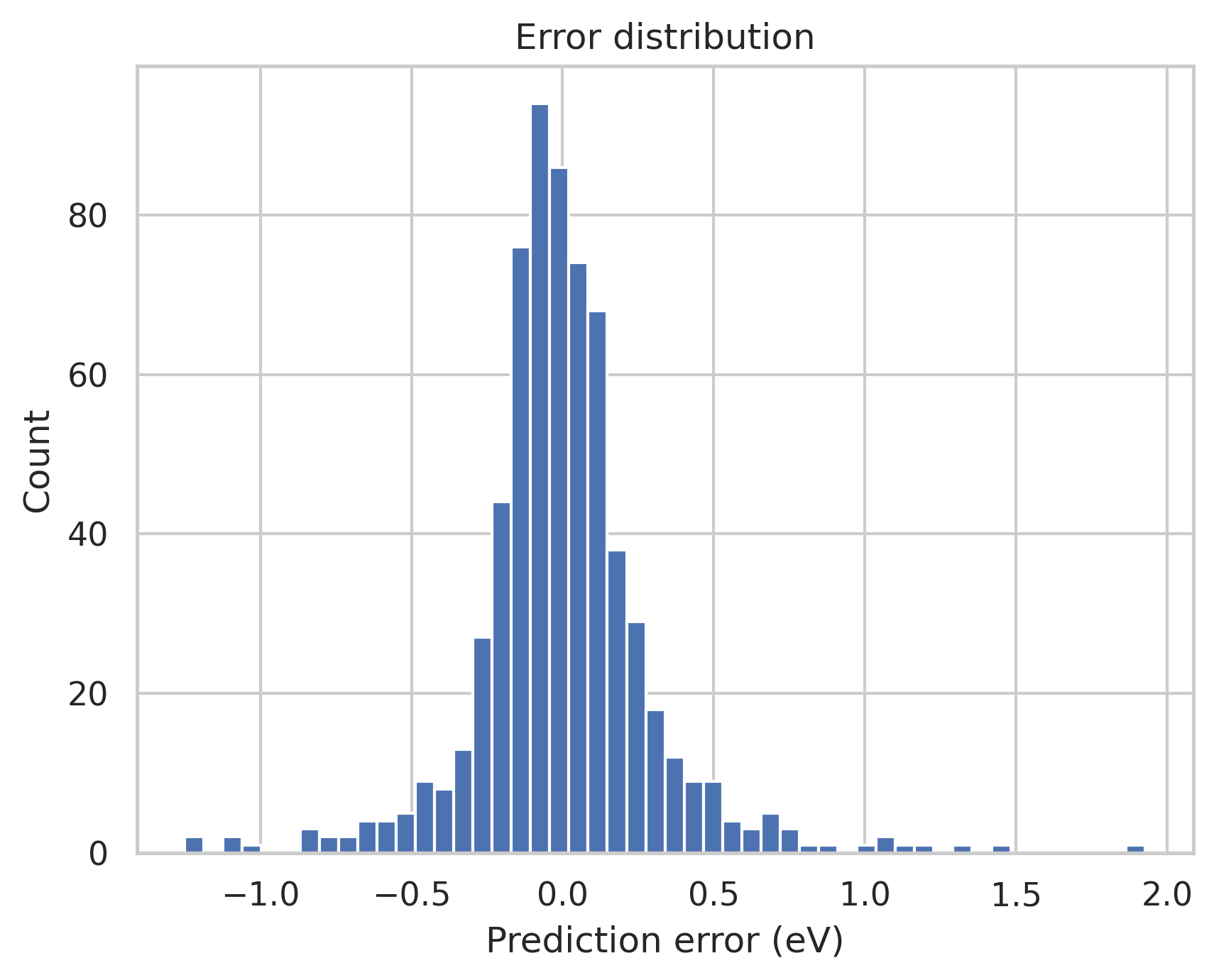}
\caption{Magpie + structural}
\end{subfigure}
\begin{subfigure}{0.32\textwidth}
\includegraphics[width=\linewidth]{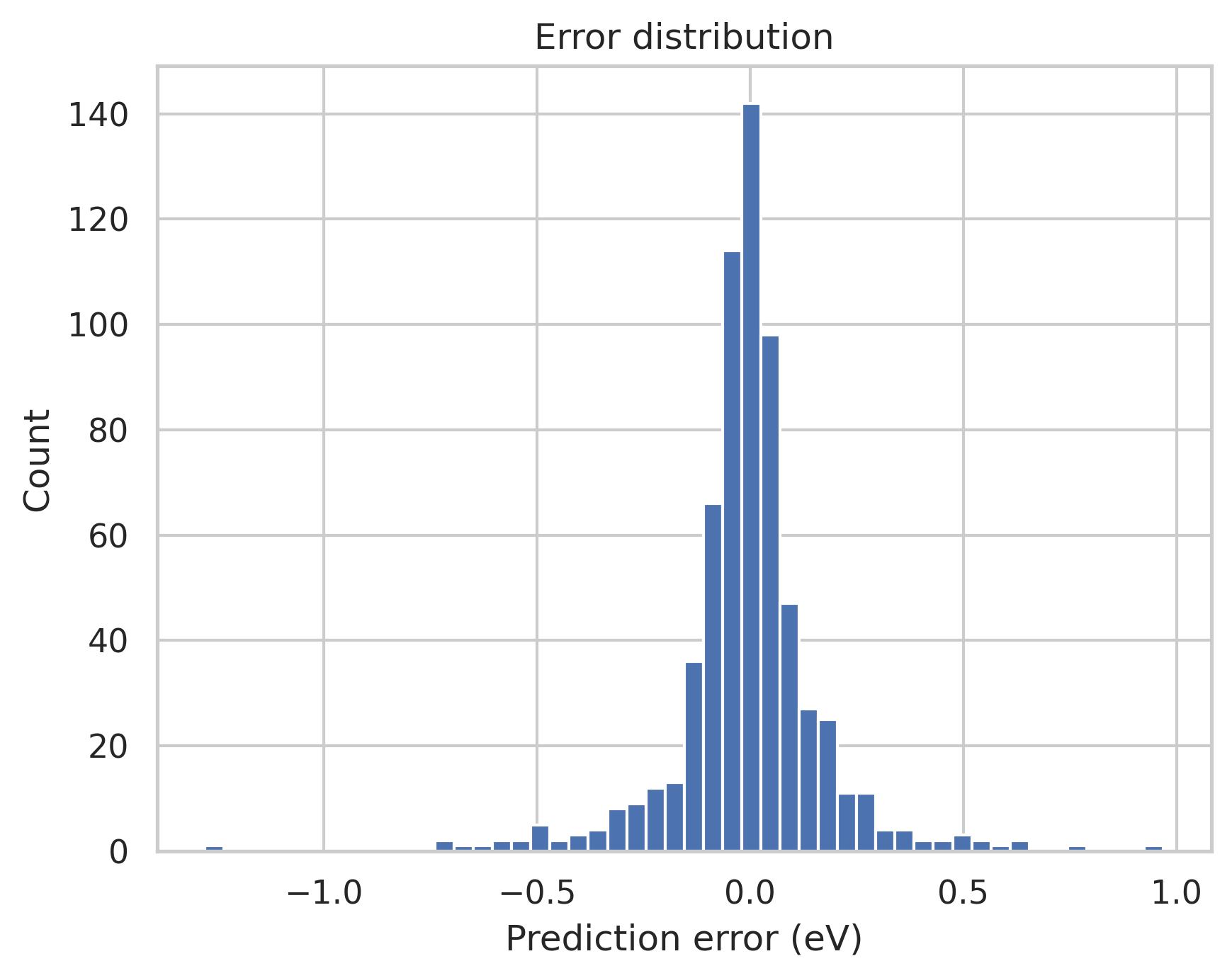}
\caption{Curated combined}
\end{subfigure}
\caption{Test-set residual distributions, defined as reconstructed minus reference HSE06 gap. The curated model has the narrowest central distribution; the remaining long negative tail corresponds to occasional under-correction of the HSE06 gap.}
\label{fig:errorhist}
\end{figure}

Residual histograms (Figure~\ref{fig:errorhist}) are centered close to zero, indicating little global signed bias. Their widths, rather than their centers, distinguish the models: the C2DB-native distribution is visibly broader, whereas the Magpie and especially curated representations concentrate more probability near zero. The occasional long negative residuals correspond to under-corrections, i.e., cases where the reconstructed gap remains smaller than the HSE06 reference. Importantly, the centered residuals do not imply that $R^2$ for $\dEg$ and the absolute gap should be equal; those $R^2$ values differ because the target variances differ, as shown in Table~\ref{tab:regression}.

\begin{figure}[!ht]
\centering
\begin{subfigure}{0.32\textwidth}
\includegraphics[width=\linewidth]{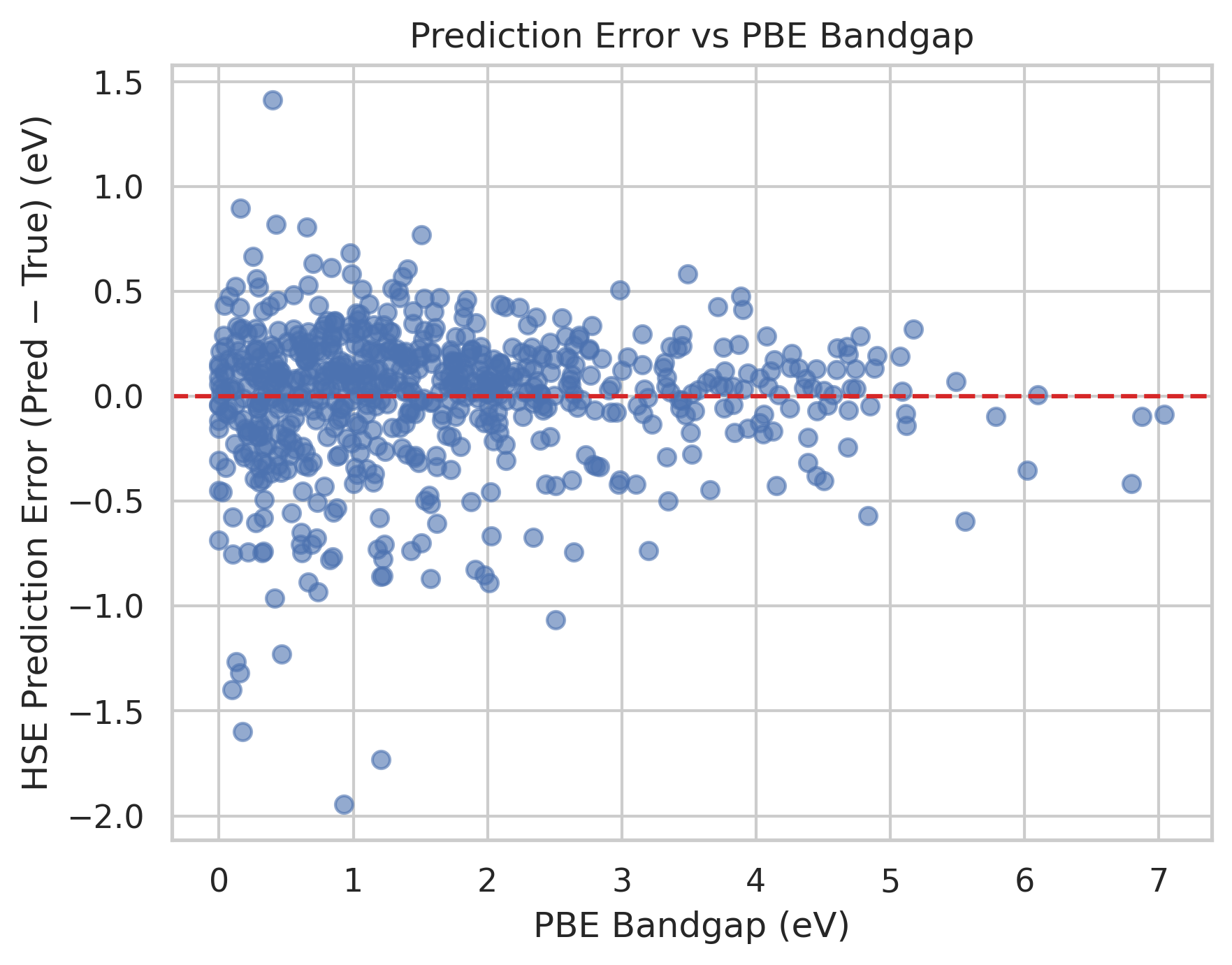}
\caption{C2DB native}
\end{subfigure}
\begin{subfigure}{0.32\textwidth}
\includegraphics[width=\linewidth]{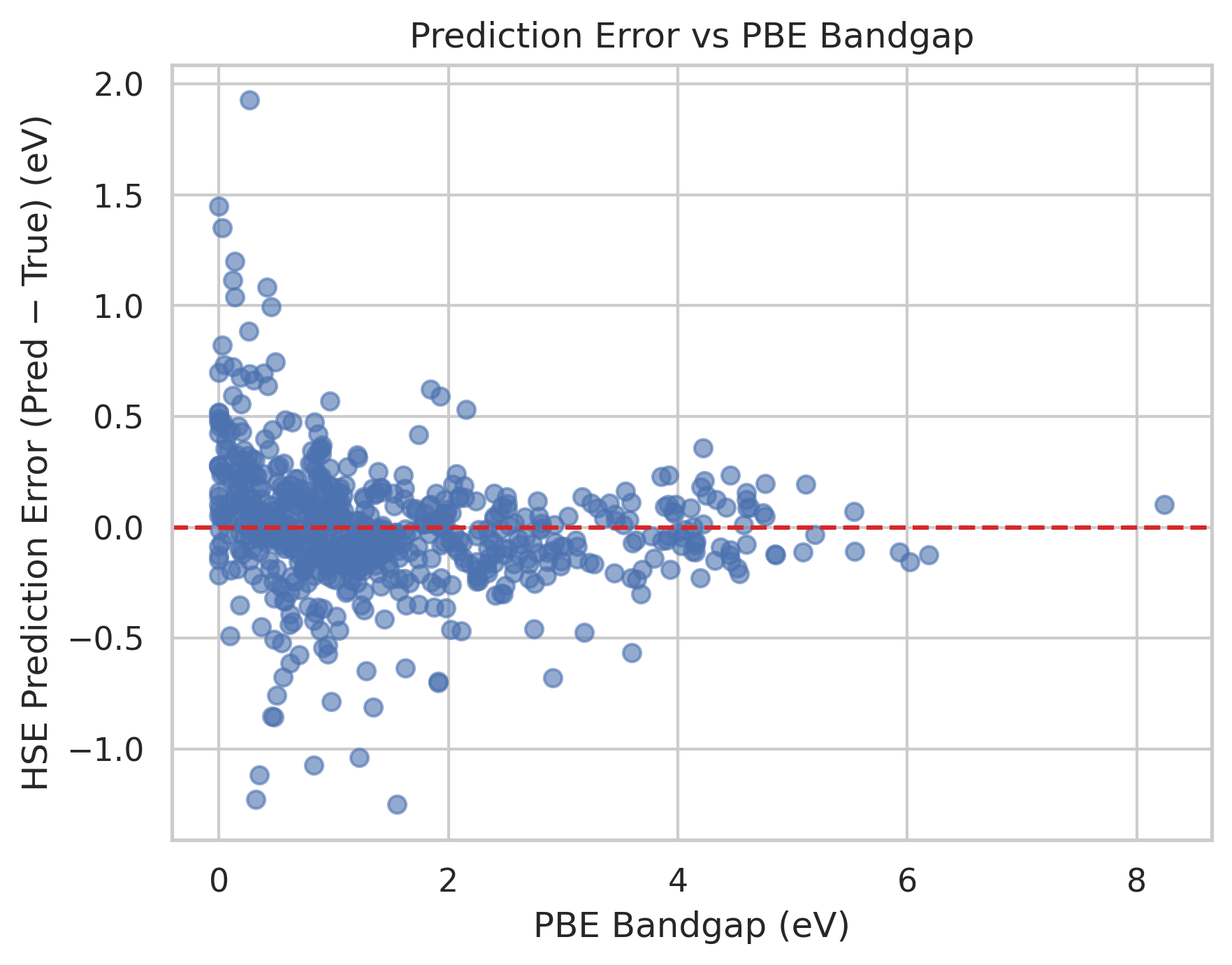}
\caption{Magpie + structural}
\end{subfigure}
\begin{subfigure}{0.32\textwidth}
\includegraphics[width=\linewidth]{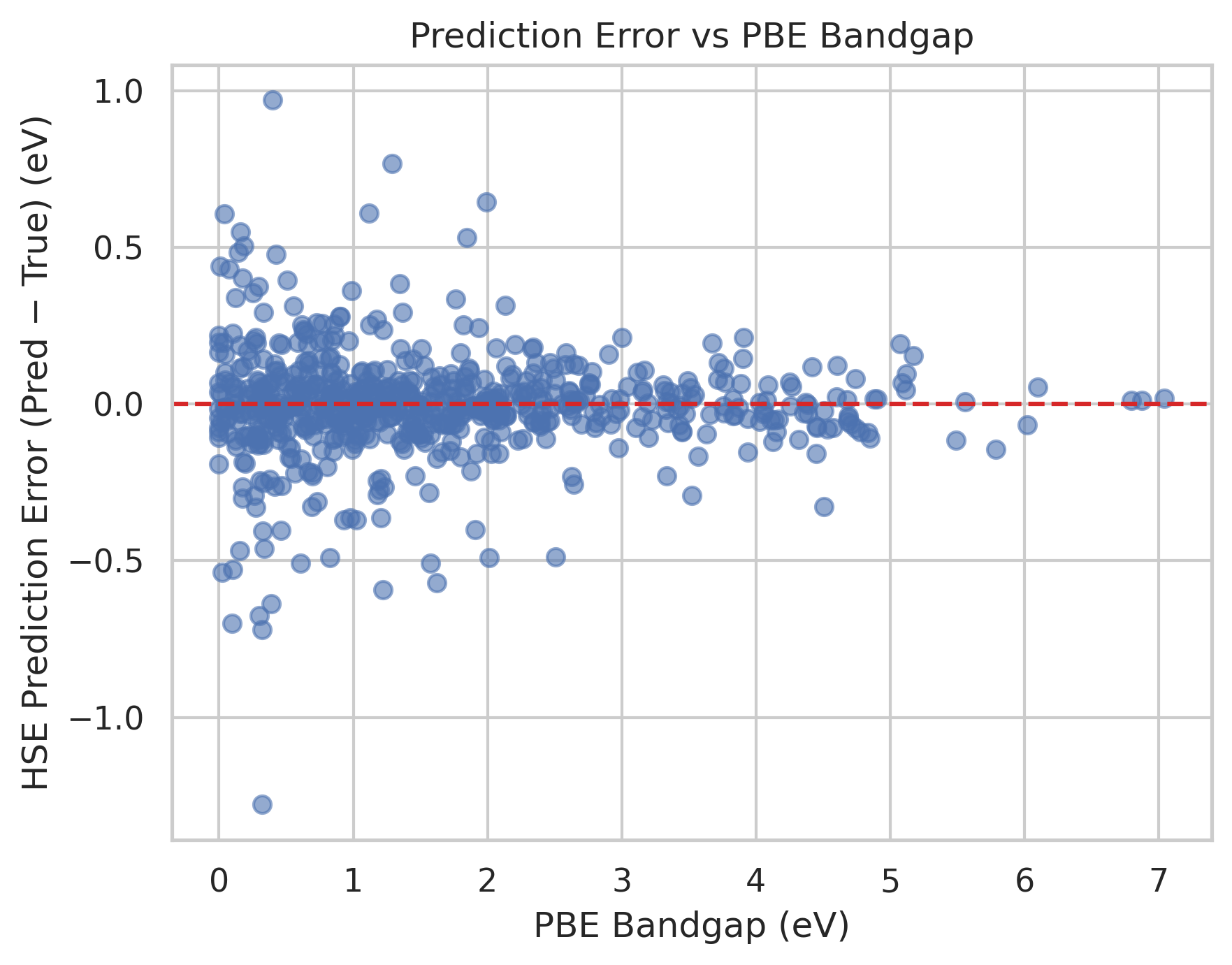}
\caption{Curated combined}
\end{subfigure}
\caption{Signed $\Delta$-learning error versus PBE gap on the regression test set. Large residuals are concentrated below approximately 1.5--2~eV, identifying the low-gap region in which explicit HSE06 validation or calibrated uncertainty estimates are most valuable for threshold-based screening.}
\label{fig:errorgap}
\end{figure}

The residual magnitude is strongly gap dependent (Figure~\ref{fig:errorgap}; absolute errors are shown in Figure~S5). Errors exceeding 0.5~eV occur predominantly for $\EgPBE\lesssim1.5$--2~eV, whereas the residual envelope contracts substantially for large PBE gaps. This is consistent with the low-gap region containing both larger and more chemistry-dependent semilocal-to-hybrid corrections. It also identifies the part of descriptor space in which uncertainty estimates would be most valuable. For screening decisions close to a band-gap threshold, point predictions should therefore be supplemented by explicit HSE06 calculations or a calibrated uncertainty model.

\FloatBarrier

\subsection{The HSE06 correction changes candidate membership, not only ranking}
A low regression MAE is useful only if it changes a materials decision. We therefore ask whether the HSE06 correction alters which compounds survive the solar-hydrogen gap window. Applying the screening criteria to the 8,221-material corrected insulator pool initially yields 10 strict ($E_{\mathrm{hull}}\le0.05$~eV/atom) and 29 relaxed ($E_{\mathrm{hull}}\le0.10$~eV/atom) candidates. The strict shortlist is given in Table~\ref{tab:candidates}; 
The 10 strict materials are nonmagnetic and have corrected gaps of 1.984--2.798~eV; eight use explicit C2DB HSE06 gaps and two (\feat{4LiSeO2-1}, \feat{4NaSeO2-1}) remain ML-derived.

\begin{figure}[!ht]
\centering
\includegraphics[width=0.98\linewidth]{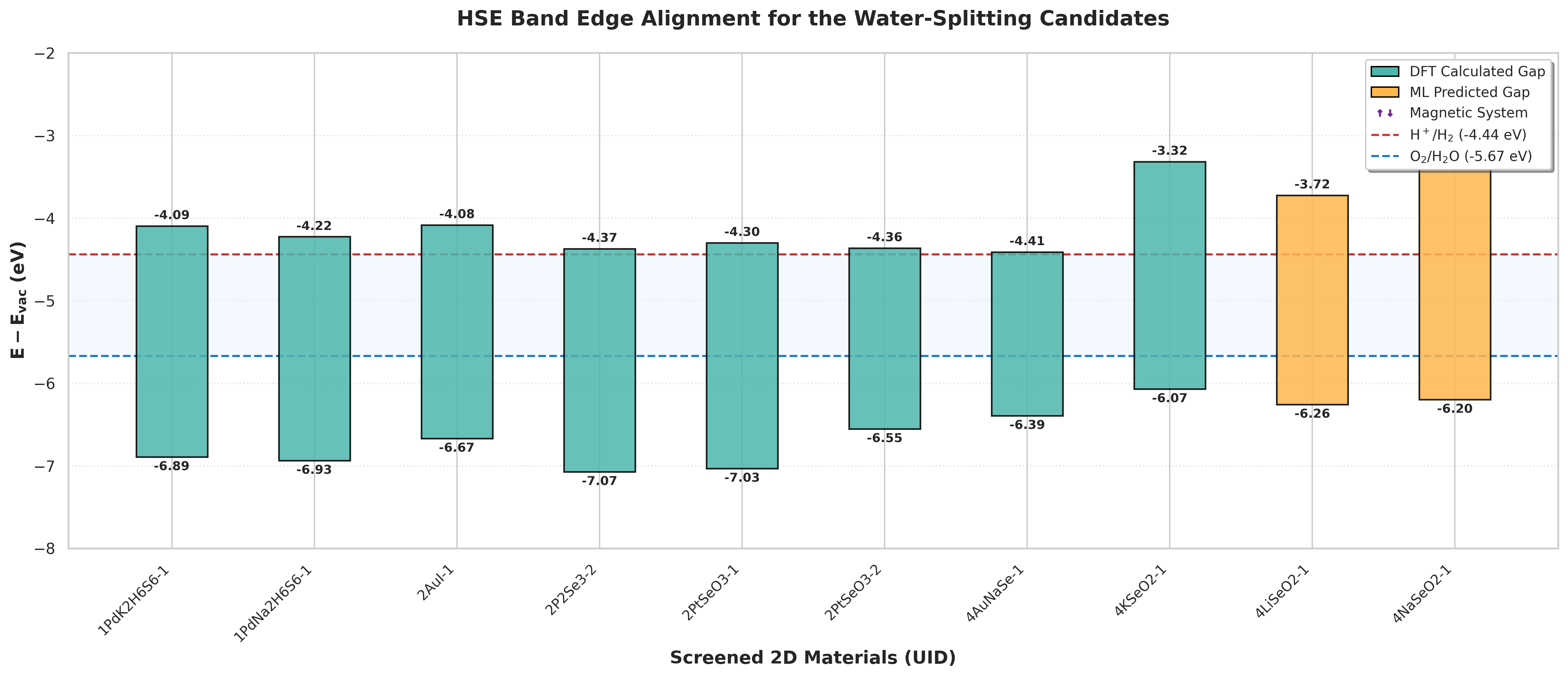}
\caption{Vacuum-aligned HSE-equivalent band edges for the strict water-splitting candidates. Teal bars denote candidates with explicit HSE06 gaps and orange bars denote candidates whose HSE06-equivalent gaps are obtained from the curated $\Delta$-learning model. Dashed horizontal lines mark the pH~0 H$^+$/H$_2$ and O$_2$/H$_2$ redox levels.}
\label{fig:bandalign}
\end{figure}

The functional correction materially changes the screen. For the strict set, $\dEg=0.592$--1.427~eV (mean 0.994~eV), and four materials would fail the same 1.6--2.8~eV gap window at PBE. In the initial relaxed set, 18 of 29 materials likewise have PBE gaps outside the window. The targeted validation then provides a second, independent effect: the explicit HSE06 gap of \feat{1AgBr-1} is 2.97~eV, above the upper cutoff, whereas \feat{1AgI-1} remains inside at 2.70~eV. The currently retained relaxed set therefore contains 28 materials (19 with explicit HSE06 gaps, including the new \feat{1AgI-1} result, and 9 still ML-derived); 17 of these 28 would lie outside the selected gap window at the PBE level. The strict set is unchanged. The resulting HSE-equivalent band-edge alignment of the strict shortlist is shown in Figure~\ref{fig:bandalign}.

\begin{table}[!ht]
\small
\centering
\caption{Strict near-hull ($E_{\mathrm{hull}}\le0.05$~eV/atom) candidates surviving the electronic water-splitting pre-screen. $\dEg=\EgHSE-\EgPBE$. ``DFT'' denotes an explicit C2DB HSE06 gap; ``ML'' denotes the curated $\Delta$-learning prediction.}
\label{tab:candidates}
\resizebox{\textwidth}{!}{%
\begin{tabular}{lcccccc}
\toprule
UID & $E_{\mathrm{hull}}$ (eV/atom) & $\Delta H_f$ (eV/atom) & $\EgPBE$ (eV) & $\EgHSE$ (eV) & $\dEg$ (eV) & Source \\
\midrule
\feat{4AuNaSe-1} & 0.011 & $-0.582$ & 1.392 & 1.984 & 0.592 & DFT \\
\feat{4KSeO2-1} & 0.000 & $-1.263$ & 1.929 & 2.751 & 0.822 & DFT \\
\feat{4LiSeO2-1} & 0.000 & $-1.279$ & 1.452 & 2.522 & 1.070 & ML \\
\feat{4NaSeO2-1} & 0.008 & $-1.234$ & 1.757 & 2.730 & 0.973 & ML \\
\feat{2PtSeO3-1} & 0.020 & $-0.590$ & 1.305 & 2.732 & 1.427 & DFT \\
\feat{2PtSeO3-2} & 0.035 & $-0.575$ & 0.828 & 2.187 & 1.359 & DFT \\
\feat{2AuI-1} & 0.000 & $-0.098$ & 1.808 & 2.585 & 0.777 & DFT \\
\feat{2P2Se3-2} & 0.028 & $-0.038$ & 1.794 & 2.700 & 0.906 & DFT \\
\feat{1PdK2H6S6-1} & 0.026 & $-0.428$ & 1.783 & 2.798 & 1.015 & DFT \\
\feat{1PdNa2H6S6-1} & 0.033 & $-0.384$ & 1.715 & 2.710 & 0.995 & DFT \\
\bottomrule
\end{tabular}%
}
\end{table}
The strict list is chemically nonuniform: seven of the ten entries contain Se, while the remaining three comprise an iodide and two Pd--S/H-containing compositions. Three strict candidates lie on the reported convex hull ($E_{\mathrm{hull}}=0$), and the remaining seven are within 35~meV/atom of it; all have negative formation enthalpies in the C2DB data. This concentration should not be interpreted as a universal chemical design rule, because it reflects both the composition of the database and the specific electronic/stability filters used here.

Before targeted validation, 18 of the 29 relaxed gaps were explicit HSE06 values and 11 were ML-derived. The new HSE06 calculations convert \feat{1AgI-1} and \feat{1AgBr-1} from ML-only cases into explicit references; \feat{1AgI-1} is retained while \feat{1AgBr-1} is removed by the gap criterion. Among the 28 currently retained relaxed candidates, 19 therefore have explicit HSE06 gaps and 9 remain ML-derived. The largest correction in this retained set is \feat{2NiTeO3-1}, whose gap increases from 0.217~eV at PBE to 2.222~eV at HSE06 ($\dEg=2.005$~eV). This example, together with the threshold crossing of \feat{1AgBr-1}, shows why high-fidelity validation must be coupled to the screening decision rather than treated only as a global error metric.

\subsection{Scope of the electronic shortlist and validation priorities}
The screen establishes three quantities only: a high-fidelity-equivalent gap, near-hull thermodynamic accessibility, and approximate pH~0 band-edge straddling. It does not establish optical absorption strength, excitonic effects, carrier transport/recombination, catalytic overpotentials, aqueous stability, photochemical corrosion resistance, or finite-temperature robustness. These are standard subsequent filters in more complete photocatalyst workflows.\cite{wang2024,shobeyrian2025,yadav2025,tang2026nanoenergy}

The highest-priority next step is direct electronic validation: explicit HSE06 gaps for the remaining ML-derived candidates and vacuum-aligned HSE06 VBM/CBM values for the strict shortlist. Because the present symmetric scissor construction corrects the gap but not the individual band-edge shifts, a gap model alone cannot quantify the uncertainty of redox straddling. Optical, catalytic, and stability calculations should then be concentrated on the candidates that survive this direct band-edge test.

\FloatBarrier

\section{Conclusions}
We developed a leakage-aware two-stage workflow in which PBE-spurious metallicity is addressed before HSE06--PBE $\Delta$-learning. The distinction between the stages is essential: classification deliberately excludes low-fidelity electronic shortcuts, whereas regression reintroduces PBE electronic quantities as the baseline to be corrected. Stage~I remains limited by the 29-material true-metal minority class (best true-metal recall 0.286), while Stage~II is supported by 3,318 HSE06-insulator references and reaches a reconstructed-gap MAE of 0.108~eV ($R^2=0.989$) with the curated XGBoost model, compared with 1.036~eV for raw PBE.

The correction has a direct materials consequence. The initial electronic pre-screen identifies 10 strict and 29 relaxed candidates, and four strict plus 18 relaxed candidates would fall outside the same visible-gap window at the PBE level. Targeted HSE06 calculations for six ML-predicted compounds provide a material-level validation set: the C2DB-native, Magpie+structural, and curated models have MAEs of 0.417, 0.182, and 0.110~eV against these new references, while \feat{1AgBr-1} demonstrates the importance of threshold-aware validation by moving above the 2.8~eV cutoff despite a modest prediction error. The current relaxed set therefore contains 28 retained candidates; the 10-material strict set is unchanged.

The remaining uncertainty is dominated by physics not yet modeled explicitly, especially individual HSE06 band-edge shifts. The shortlist should therefore be viewed as a prioritized validation set rather than as a claim of established photocatalytic performance. More generally, the study shows that low-to-high-fidelity learning is most useful when coupled to upstream population repair and downstream decision criteria, so that model accuracy is evaluated in terms of which materials are actually admitted to or removed from an energy-materials screen.

\begin{suppinfo}
Additional classification diagnostics (feature correlations, cross-validation, all-model and training-set confusion matrices); additional regression diagnostics (feature-importance comparisons, absolute-error dependence on PBE gap, and the complete material-resolved HSE06 validation table) are given in the Supporting Information (SI) file.
\end{suppinfo}

\bibliography{refer}

@article{wang2020domen,
  author  = {Wang, Qian and Domen, Kazunari},
  title   = {Particulate Photocatalysts for Light-Driven Water Splitting: Mechanisms, Challenges, and Design Strategies},
  journal = {Chem. Rev.},
  year    = {2020},
  volume  = {120},
  pages   = {919--985},
  doi     = {10.1021/acs.chemrev.9b00201}
}

@article{mai2022chemrev,
  author  = {Mai, Haoxin and Le, Tu C. and Chen, Dehong and Winkler, David A. and Caruso, Rachel A.},
  title   = {Machine Learning for Electrocatalyst and Photocatalyst Design and Discovery},
  journal = {Chem. Rev.},
  year    = {2022},
  volume  = {122},
  pages   = {13478--13515},
  doi     = {10.1021/acs.chemrev.2c00061}
}

@article{jin2022jpcl,
  author  = {Jin, Hao and Tan, Xiaoxing and Wang, Tao and Yu, Yunjin and Wei, Yadong},
  title   = {Discovery of Two-Dimensional Multinary Component Photocatalysts Accelerated by Machine Learning},
  journal = {J. Phys. Chem. Lett.},
  year    = {2022},
  volume  = {13},
  pages   = {7228--7235},
  doi     = {10.1021/acs.jpclett.2c01862}
}

@article{gao2024acsnano,
  author  = {Gao, Yunzhi and Zhang, Qian and Hu, Wei and Yang, Jinlong},
  title   = {First-Principles Computational Screening of Two-Dimensional Polar Materials for Photocatalytic Water Splitting},
  journal = {ACS Nano},
  year    = {2024},
  volume  = {18},
  pages   = {19381--19390},
  doi     = {10.1021/acsnano.4c06544}
}

@article{haastrup2018,
  author  = {Haastrup, Sten and Strange, Mikkel and Pandey, Mohnish and Deilmann, Thorsten and Schmidt, Per S. and Hinsche, Nicki F. and Gjerding, Morten N. and Torelli, Daniele and Larsen, Peter M. and Riis-Jensen, Anders C. and Gath, Jakob and Jacobsen, Karsten W. and Mortensen, Jens J{\o}rgen and Olsen, Thomas and Thygesen, Kristian S.},
  title   = {The Computational 2D Materials Database: High-Throughput Modeling and Discovery of Atomically Thin Crystals},
  journal = {2D Mater.},
  year    = {2018},
  volume  = {5},
  pages   = {042002},
  doi     = {10.1088/2053-1583/aacfc1}
}

@article{gjerding2021,
  author  = {Gjerding, Morten Niklas and Taghizadeh, Alireza and Rasmussen, Asbj{\o}rn and Ali, Sajid and Bertoldo, Fabian and Deilmann, Thorsten and Holguin, Urko Petralanda and Kn{\o}sgaard, Nikolaj R{\o}rb{\ae}k and Kruse, Mads and Larsen, Ask Hjorth and Manti, Simone and Pedersen, Thomas Garm and Skovhus, Thorbj{\o}rn and Svendsen, Mark Kamper and Mortensen, Jens J{\o}rgen and Olsen, Thomas and Thygesen, Kristian Sommer},
  title   = {Recent Progress of the Computational 2D Materials Database (C2DB)},
  journal = {2D Mater.},
  year    = {2021},
  volume  = {8},
  pages   = {044002},
  doi     = {10.1088/2053-1583/ac1059}
}

@article{singh2015,
  author  = {Singh, Arunima K. and Mathew, Kiran and Zhuang, Houlong L. and Hennig, Richard G.},
  title   = {Computational Screening of 2D Materials for Photocatalysis},
  journal = {J. Phys. Chem. Lett.},
  year    = {2015},
  volume  = {6},
  pages   = {1087--1098},
  doi     = {10.1021/jz502646d}
}

@article{wang2024jpcl,
  author  = {Wang, Yatong and Sorkun, Murat Cihan and Brocks, Geert and Er, S{\"u}leyman},
  title   = {ML-Aided Computational Screening of 2D Materials for Photocatalytic Water Splitting},
  journal = {J. Phys. Chem. Lett.},
  year    = {2024},
  volume  = {15},
  pages   = {4983--4991},
  doi     = {10.1021/acs.jpclett.4c00425}
}

@article{wang2024,
  author  = {Wang, Yatong and Brocks, Geert and Er, S{\"u}leyman},
  title   = {Data-Driven Discovery of Intrinsic Direct-Gap 2D Materials as Potential Photocatalysts for Efficient Water Splitting},
  journal = {ACS Catal.},
  year    = {2024},
  volume  = {14},
  pages   = {1336--1350},
  doi     = {10.1021/acscatal.3c05181}
}

@article{pbe,
  author  = {Perdew, John P. and Burke, Kieron and Ernzerhof, Matthias},
  title   = {Generalized Gradient Approximation Made Simple},
  journal = {Phys. Rev. Lett.},
  year    = {1996},
  volume  = {77},
  pages   = {3865--3868},
  doi     = {10.1103/PhysRevLett.77.3865}
}

@article{kresse1996prb,
  author  = {Kresse, Georg and Furthm{\"u}ller, J{\"u}rgen},
  title   = {Efficient Iterative Schemes for {\textit{ab initio}} Total-Energy Calculations Using a Plane-Wave Basis Set},
  journal = {Phys. Rev. B},
  year    = {1996},
  volume  = {54},
  pages   = {11169--11186},
  doi     = {10.1103/PhysRevB.54.11169}
}

@article{kresse1996cms,
  author  = {Kresse, Georg and Furthm{\"u}ller, J{\"u}rgen},
  title   = {Efficiency of {\textit{ab initio}} Total Energy Calculations for Metals and Semiconductors Using a Plane-Wave Basis Set},
  journal = {Comput. Mater. Sci.},
  year    = {1996},
  volume  = {6},
  pages   = {15--50},
  doi     = {10.1016/0927-0256(96)00008-0}
}

@article{bloechl1994,
  author  = {Bl{\"o}chl, Peter E.},
  title   = {Projector Augmented-Wave Method},
  journal = {Phys. Rev. B},
  year    = {1994},
  volume  = {50},
  pages   = {17953--17979},
  doi     = {10.1103/PhysRevB.50.17953}
}

@article{kresse1999,
  author  = {Kresse, Georg and Joubert, Daniel},
  title   = {From Ultrasoft Pseudopotentials to the Projector Augmented-Wave Method},
  journal = {Phys. Rev. B},
  year    = {1999},
  volume  = {59},
  pages   = {1758--1775},
  doi     = {10.1103/PhysRevB.59.1758}
}

@article{heyd2003,
  author  = {Heyd, Jochen and Scuseria, Gustavo E. and Ernzerhof, Matthias},
  title   = {Hybrid Functionals Based on a Screened Coulomb Potential},
  journal = {J. Chem. Phys.},
  year    = {2003},
  volume  = {118},
  pages   = {8207--8215},
  doi     = {10.1063/1.1564060}
}

@article{krukau2006,
  author  = {Krukau, Aliaksandr V. and Vydrov, Oleg A. and Izmaylov, Artur F. and Scuseria, Gustavo E.},
  title   = {Influence of the Exchange Screening Parameter on the Performance of Screened Hybrid Functionals},
  journal = {J. Chem. Phys.},
  year    = {2006},
  volume  = {125},
  pages   = {224106},
  doi     = {10.1063/1.2404663}
}

@article{ramakrishnan2015,
  author  = {Ramakrishnan, Raghunathan and Dral, Pavlo O. and Rupp, Matthias and von Lilienfeld, O. Anatole},
  title   = {Big Data Meets Quantum Chemistry Approximations: The $\Delta$-Machine Learning Approach},
  journal = {J. Chem. Theory Comput.},
  year    = {2015},
  volume  = {11},
  pages   = {2087--2096},
  doi     = {10.1021/acs.jctc.5b00099}
}

@article{adhikari2023,
  author  = {Adhikari, Santosh and Clary, Jacob and Sundararaman, Ravishankar and Musgrave, Charles B. and Vigil-Fowler, Derek and Sutton, Christopher},
  title   = {Accurate Prediction of HSE06 Band Structures for a Diverse Set of Materials Using $\Delta$-Learning},
  journal = {Chem. Mater.},
  year    = {2023},
  volume  = {35},
  number  = {20},
  pages   = {8397--8405},
  doi     = {10.1021/acs.chemmater.3c01131}
}

@article{karimitari2025,
  author  = {Karimitari, Nima and Pakornchote, Teerachote and Alherz, Abdulaziz W. and Clary, Jacob M. and Tezak, Cooper and Dey, Sourin and Hu, Jianjun and Vigil-Fowler, Derek and Sundararaman, Ravishankar and Musgrave, Charles B. and Sutton, Christopher},
  title   = {$\Delta$-Learning of High-Fidelity Electronic Structure Using Graph Neural Networks with Modified Node-Level Features},
  journal = {ACS Mater. Lett.},
  year    = {2025},
  volume  = {7},
  number  = {12},
  pages   = {3901--3907},
  doi     = {10.1021/acsmaterialslett.5c00857}
}

@article{kang2026orbnet,
  author  = {Kang, Minhyuk and Kang, Beom Seok and Kliavinek, Sergei and Hanisch, Maurice D. and Goddard, William A. and Anandkumar, Anima},
  title   = {An Orbital-Based Geometric Deep Learning Framework for Periodic Materials},
  journal = {International Conference on Learning Representations (ICLR)},
  year    = {2026},
  note    = {\url{https://openreview.net/forum?id=Dnjz9nlGBq}}
}

@article{tang2026nanoenergy,
  author  = {Tang, Kan and Liu, Shengxian and Zhou, Jian and Sun, Zhimei},
  title   = {Data-Driven Discovery of Highly Efficient 2D Photocatalytic Materials for Water Splitting},
  journal = {Nano Energy},
  year    = {2026},
  volume  = {153},
  pages   = {111929},
  doi     = {10.1016/j.nanoen.2026.111929}
}

@article{kapoor2023,
  author  = {Kapoor, Sayash and Narayanan, Arvind},
  title   = {Leakage and the Reproducibility Crisis in Machine-Learning-Based Science},
  journal = {Patterns},
  year    = {2023},
  volume  = {4},
  pages   = {100804},
  doi     = {10.1016/j.patter.2023.100804}
}

@article{georgescu2021,
  author  = {Georgescu, Alexandru B. and Ren, Peiwen and Toland, Alycia R. and Zhang, Shuyi and Miller, Kyle D. and Apley, Daniel W. and Olivetti, Elsa A. and Wagner, Nikolas and Rondinelli, James M.},
  title   = {Database, Features, and Machine Learning Model to Identify Thermally Driven Metal--Insulator Transition Compounds},
  journal = {Chem. Mater.},
  year    = {2021},
  volume  = {33},
  pages   = {5591--5605},
  doi     = {10.1021/acs.chemmater.1c00905}
}

@article{ward2016,
  author  = {Ward, Logan and Agrawal, Ankit and Choudhary, Alok and Wolverton, Christopher},
  title   = {A General-Purpose Machine Learning Framework for Predicting Properties of Inorganic Materials},
  journal = {npj Comput. Mater.},
  year    = {2016},
  volume  = {2},
  pages   = {16028},
  doi     = {10.1038/npjcompumats.2016.28}
}

@article{ward2018,
  author  = {Ward, Logan and Dunn, Alexander and Faghaninia, Alireza and Zimmermann, Nils E. R. and Bajaj, Saurabh and Wang, Qi and Montoya, Joseph and Chen, Jiming and Bystrom, Kyle and Dylla, Maxwell and Chard, Kyle and Asta, Mark and Persson, Kristin A. and Snyder, G. Jeffrey and Foster, Ian and Jain, Anubhav},
  title   = {Matminer: An Open Source Toolkit for Materials Data Mining},
  journal = {Comput. Mater. Sci.},
  year    = {2018},
  volume  = {152},
  pages   = {60--69},
  doi     = {10.1016/j.commatsci.2018.05.018}
}

@article{ong2013,
  author  = {Ong, Shyue Ping and Richards, William Davidson and Jain, Anubhav and Hautier, Geoffroy and Kocher, Michael and Cholia, Shreyas and Gunter, Dan and Chevrier, Vincent L. and Persson, Kristin A. and Ceder, Gerbrand},
  title   = {Python Materials Genomics (pymatgen): A Robust, Open-Source Python Library for Materials Analysis},
  journal = {Comput. Mater. Sci.},
  year    = {2013},
  volume  = {68},
  pages   = {314--319},
  doi     = {10.1016/j.commatsci.2012.10.028}
}

@article{larsen2017,
  author  = {Larsen, Ask Hjorth and Mortensen, Jens J{\o}rgen and Blomqvist, Jakob and Castelli, Ivano E. and Christensen, Rune and Du{\l}ak, Marcin and Friis, Jesper and Groves, Michael N. and Hammer, Bj{\o}rk and Hargus, Cory and Hermes, Eric D. and Jennings, Paul C. and Jensen, Peter Bjerre and Kermode, James and Kitchin, John R. and Kolsbjerg, Esben Leonhard and Kubal, Joseph and Kaasbjerg, Kristen and Lysgaard, Steen and Maronsson, J{\'o}n Bergmann and Maxson, Tristan and Olsen, Thomas and Pastewka, Lars and Peterson, Andrew and Rostgaard, Carsten and Schi{\o}tz, Jakob and Sch{\"u}tt, Ole and Strange, Mikkel and Thygesen, Kristian S. and Vegge, Tejs and Vilhelmsen, Lasse and Walter, Michael and Zeng, Zhenhua and Jacobsen, Karsten W.},
  title   = {The Atomic Simulation Environment---A Python Library for Working with Atoms},
  journal = {J. Phys.: Condens. Matter},
  year    = {2017},
  volume  = {29},
  pages   = {273002},
  doi     = {10.1088/1361-648X/aa680e}
}

@article{xgboost,
  author  = {Chen, Tianqi and Guestrin, Carlos},
  title   = {XGBoost: A Scalable Tree Boosting System},
  journal = {Proceedings of the 22nd ACM SIGKDD International Conference on Knowledge Discovery and Data Mining},
  year    = {2016},
  pages   = {785--794},
  doi     = {10.1145/2939672.2939785}
}

@article{pedregosa2011,
  author  = {Pedregosa, Fabian and Varoquaux, Ga{\"e}l and Gramfort, Alexandre and Michel, Vincent and Thirion, Bertrand and Grisel, Olivier and Blondel, Mathieu and Prettenhofer, Peter and Weiss, Ron and Dubourg, Vincent and Vanderplas, Jake and Passos, Alexandre and Cournapeau, David and Brucher, Matthieu and Perrot, Matthieu and Duchesnay, {\'E}douard},
  title   = {Scikit-learn: Machine Learning in Python},
  journal = {J. Mach. Learn. Res.},
  year    = {2011},
  volume  = {12},
  pages   = {2825--2830}
}

@article{butler1978,
  author  = {Butler, M. A. and Ginley, D. S.},
  title   = {Prediction of Flatband Potentials at Semiconductor--Electrolyte Interfaces from Atomic Electronegativities},
  journal = {J. Electrochem. Soc.},
  year    = {1978},
  volume  = {125},
  pages   = {228--232},
  doi     = {10.1149/1.2131419}
}

@article{xu2000,
  author  = {Xu, Yong and Schoonen, Martin A. A.},
  title   = {The Absolute Energy Positions of Conduction and Valence Bands of Selected Semiconducting Minerals},
  journal = {Am. Mineral.},
  year    = {2000},
  volume  = {85},
  pages   = {543--556},
  doi     = {10.2138/am-2000-0416}
}

@article{shobeyrian2025,
  author  = {Shobeyrian, Forough and Soleimani, Maryam and Shojaei, Fazel and Lashani Zand, Ali and Pourfath, Mahdi},
  title   = {Computational Screening of 2D Heterostructures for Efficient Photocatalytic Water Splitting},
  journal = {ACS Appl. Energy Mater.},
  year    = {2025},
  volume  = {8},
  pages   = {9748--9759},
  doi     = {10.1021/acsaem.5c01342}
}

@article{yadav2025,
  author  = {Yadav, Shivanand and Modi, Jainandan Kumar and Ahammed, Raihan and Bhadoria, B. S. and Chauhan, Yogesh S. and Agarwal, Amit and Bhowmick, Somnath},
  title   = {High-Throughput Screening of 2D Photocatalyst Heterostructures with Suppressed Electron--Hole Recombination for Solar Water Splitting},
  journal = {ACS Appl. Energy Mater.},
  year    = {2025},
  volume  = {8},
  pages   = {14601--14611},
  doi     = {10.1021/acsaem.5c02326}
}

\section{Supplemental Material}

This Supporting Information reports diagnostics and extended results that complement the main text. The analyses use the same train/test partitions, model definitions, descriptor names, and screening thresholds as the main manuscript. Section, figure, and table numbering is prefixed by ``S'' and is independent of the main-text numbering.

\section{Additional Classification Diagnostics}

\subsection{Correlation structure of the compact C2DB-native representation}
\begin{figure}[!ht]
\centering
\includegraphics[width=0.72\linewidth]{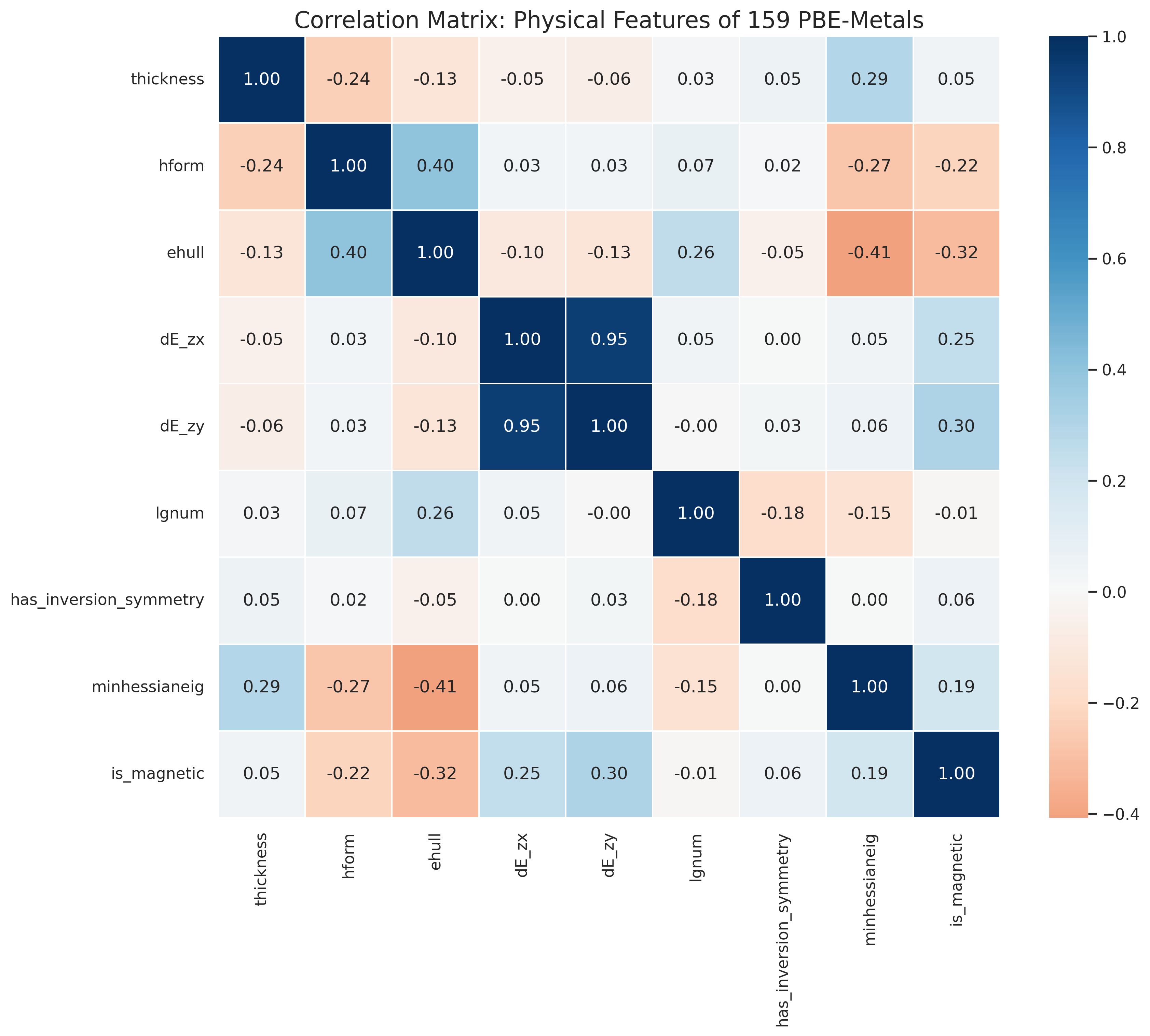}
\caption{Pairwise Pearson correlation matrix for the nine non-electronic C2DB-native descriptors used in Stage~I, evaluated over the 159-material HSE06-labeled classification pool.}
\label{figS:corr}
\end{figure}

Figure~\ref{figS:corr} shows the pairwise Pearson correlation matrix for the nine non-electronic C2DB-native descriptors used in Stage~I classification, evaluated over the 159-material HSE06-labeled PBE-metal population. Most pairs are weakly correlated ($|r|<0.3$). The strongest relationship is between the two symmetry-breaking energy descriptors \feat{dE\_zx} and \feat{dE\_zy} ($r=0.95$), which probe closely related distortion energetics. Formation enthalpy and energy above hull show a moderate positive correlation ($r=0.40$), as expected for related thermodynamic descriptors.

No correlation pruning was applied to this nine-feature representation or to the 141-feature Magpie+structural representation. Correlation pruning was restricted to the high-dimensional curated representation: 138 columns were removed from the 841-column raw table at $|r|>0.95$, leaving 703 columns before the stage-specific inclusion/exclusion rules produced 694 classification and 702 regression features.

\subsection{Cross-validation comparison and model-selection rationale}
Table~\ref{tabS:cv} reports the complete 5-fold stratified cross-validation comparison available in the final notebooks for all three descriptor representations. The values are mean $\pm$ standard deviation across folds.

For C2DB-native descriptors, the RBF-SVC has the highest mean cross-validation accuracy, but Random Forest has the highest holdout accuracy and a small train--test gap. For the Magpie+structural representation, Random Forest has the highest mean cross-validation and holdout accuracy. For the curated representation, Random Forest is close to the best mean cross-validation accuracy while retaining the strongest holdout performance. This stability, together with the use of a common model family across representations, motivated selection of Random Forest as the Stage~I classifier. The small number of minority-class examples should nevertheless be kept in mind when interpreting fold-to-fold differences.

\begin{table}[!ht]
\small
\centering
\caption{Five-fold stratified cross-validation accuracy (mean $\pm$ standard deviation) for the four candidate classifiers on the training folds of the 159-material labeled pool.}
\label{tabS:cv}
\begin{tabular}{lccc}
\toprule
Classifier & C2DB native & Magpie + structural & Curated combined \\
\midrule
SVC (RBF) & $82.4\pm1.6\%$ & $80.7\pm1.9\%$ & $84.1\pm1.5\%$ \\
Random Forest & $81.5\pm1.9\%$ & $83.2\pm2.5\%$ & $82.4\pm3.0\%$ \\
Gradient Boosting & $79.9\pm3.0\%$ & $79.9\pm5.5\%$ & $83.2\pm3.6\%$ \\
XGBoost & $80.7\pm3.4\%$ & $80.7\pm1.9\%$ & $81.5\pm1.9\%$ \\
\bottomrule
\end{tabular}
\end{table}

\subsection{Holdout confusion matrices for the selected classifier}
\begin{figure}[!ht]
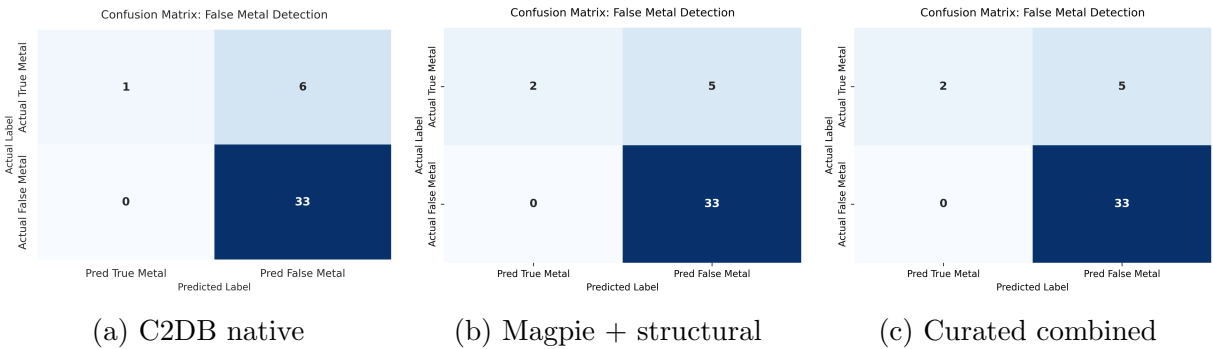

\centering
\begin{subfigure}{0.32\textwidth}
\includegraphics[width=\linewidth]{Figures/confusion-matrix_on_test-set_classification_FS-1.png}
\caption{C2DB native}
\end{subfigure}
\begin{subfigure}{0.32\textwidth}
\includegraphics[width=\linewidth]{Figures/confusion-matrix_on_test-set_classification_FS-2.png}
\caption{Magpie + structural}
\end{subfigure}
\begin{subfigure}{0.32\textwidth}
\includegraphics[width=\linewidth]{Figures/confusion-matrix_on_test-set_classification_FS-3.png}
\caption{Curated combined}
\end{subfigure}
\caption{Holdout confusion matrices for the selected Random Forest classifier using each descriptor representation.}
\label{figS:allconf}
\end{figure}

Figure~\ref{figS:allconf} reproduces the holdout confusion matrices for the selected Random Forest classifier across all three descriptor representations. The common test fold contains 33 PBE-spurious metals (HSE06 insulators) and 7 true HSE06 metals. The figure highlights why raw accuracy alone is insufficient for model selection in this imbalanced set. The main manuscript therefore reports class-resolved metrics and balanced accuracy for the selected Random Forest classifiers.

\subsection{Training-set fit of the selected Random Forest models}
\begin{figure}[!ht]
\centering
\begin{subfigure}{0.32\textwidth}
\includegraphics[width=\linewidth]{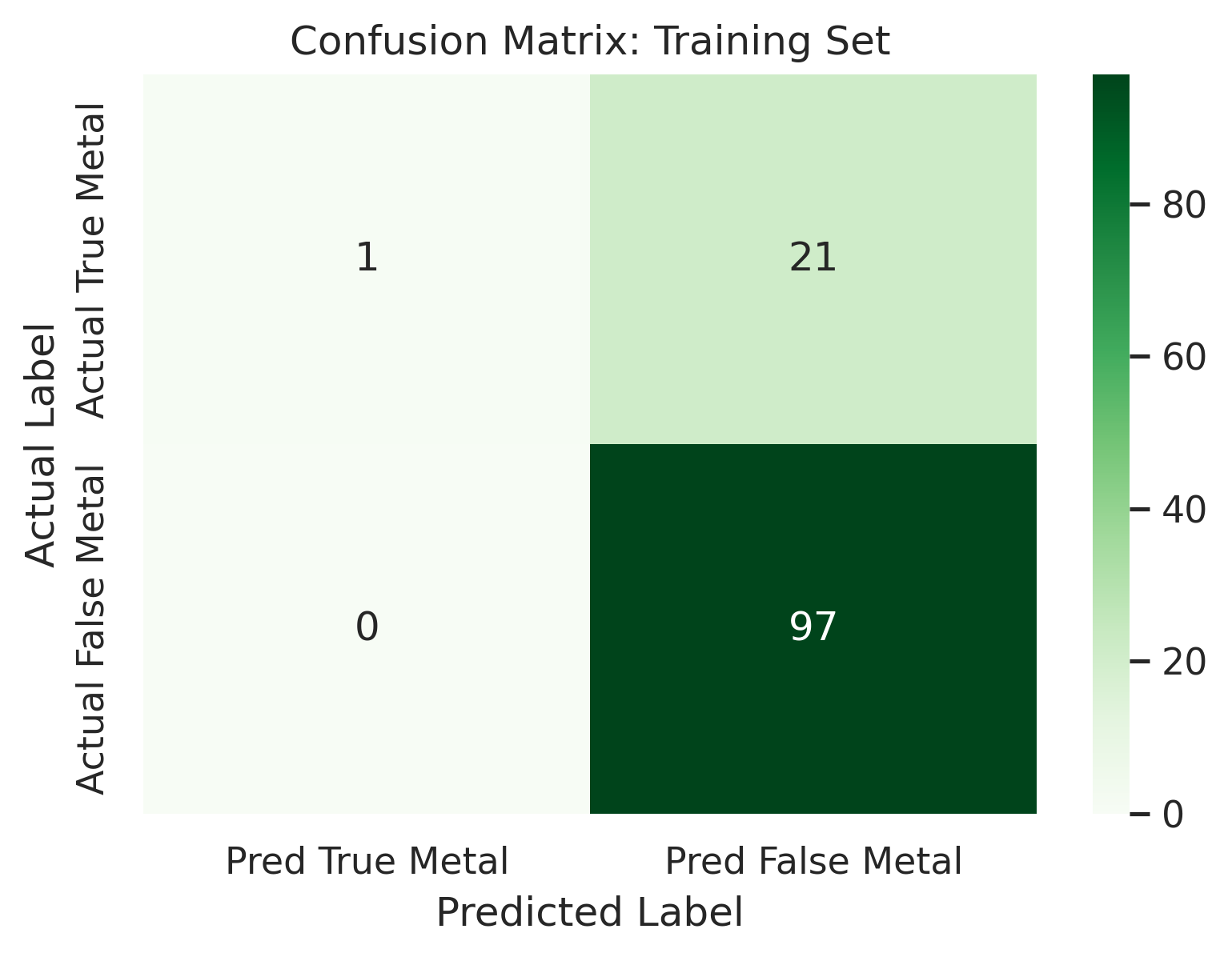}
\caption{C2DB native}
\end{subfigure}
\begin{subfigure}{0.32\textwidth}
\includegraphics[width=\linewidth]{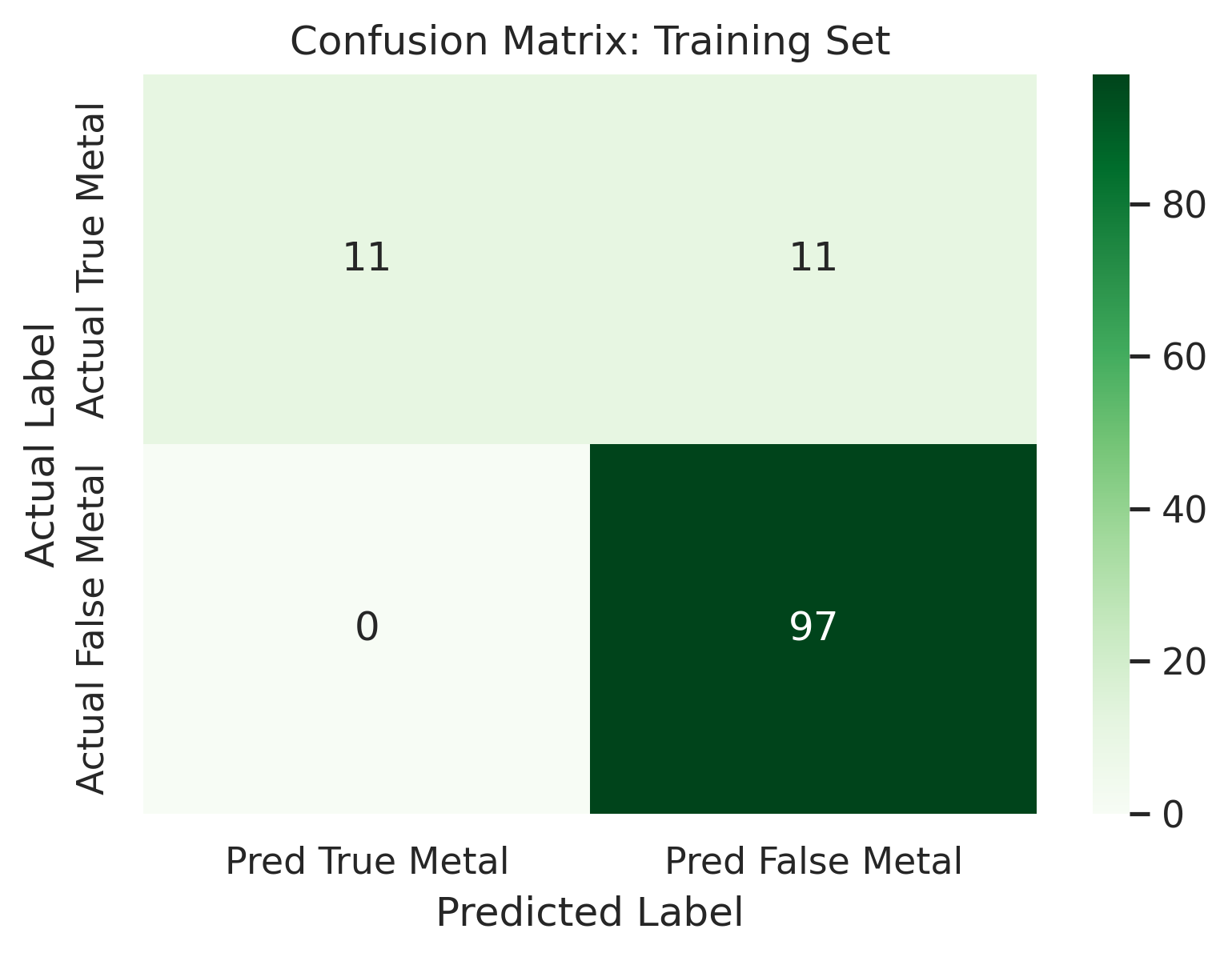}
\caption{Magpie + structural}
\end{subfigure}
\begin{subfigure}{0.32\textwidth}
\includegraphics[width=\linewidth]{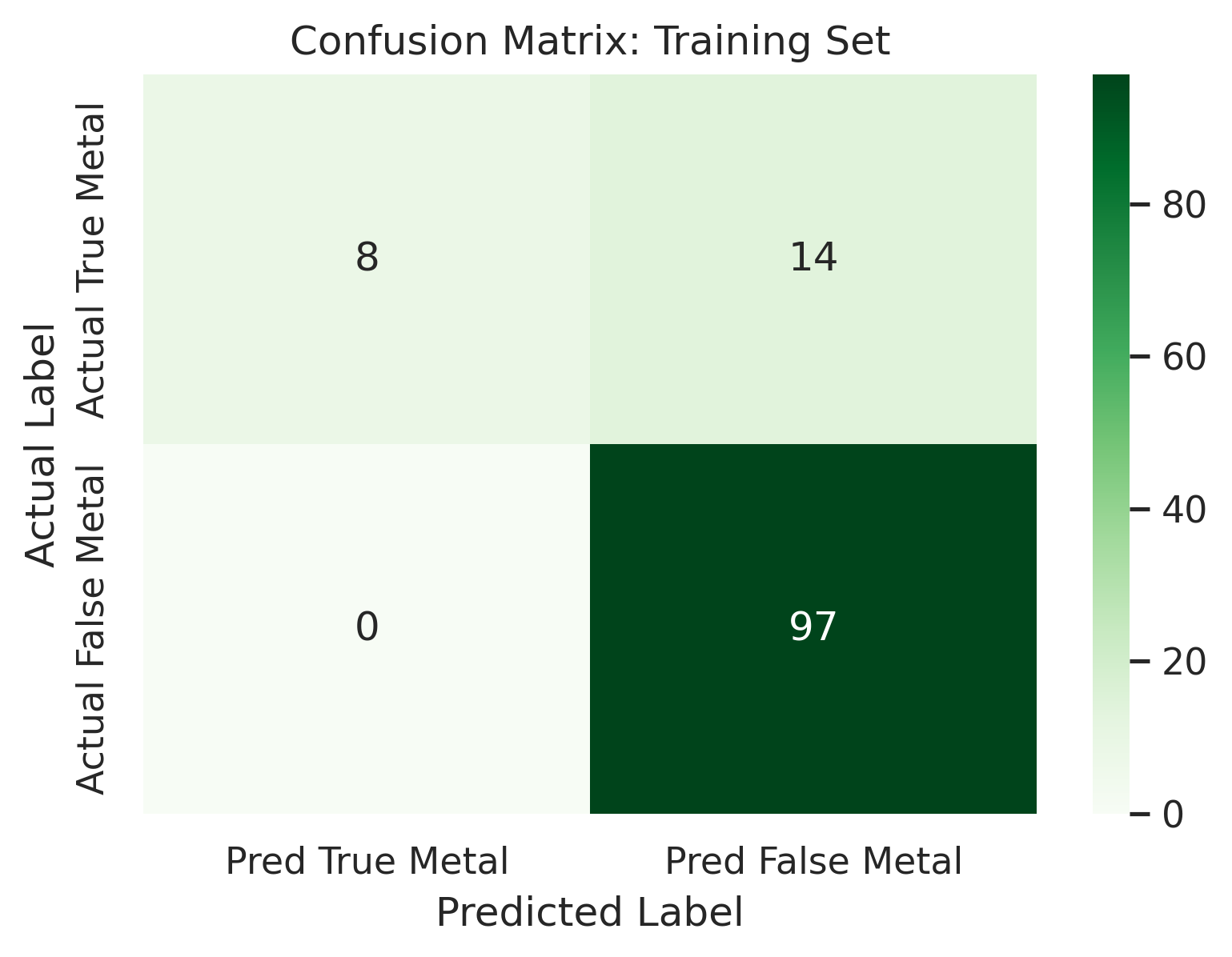}
\caption{Curated combined}
\end{subfigure}
\caption{Training-set confusion matrices for the selected Random Forest classifier (119 materials) using the three descriptor representations.}
\label{figS:trainconf}
\end{figure}

The selected Random Forest estimators nearly perfectly separate the 119 training examples (Figure~\ref{figS:trainconf}): the C2DB-native representation fits all 119 correctly, while the Magpie and curated representations each misclassify only one majority-class example. This behavior should not be described as evidence that the classification problem is intrinsically easy. A depth-5, 100-tree ensemble has sufficient capacity to fit a small data set, and the large difference between training separation and held-out minority recall indicates substantial statistical uncertainty. Consequently, the cross-validation and held-out class-resolved metrics are more informative than training accuracy.

\FloatBarrier

\section{Additional Regression Diagnostics}

\subsection{Feature importance for C2DB-native and Magpie $\Delta$-learning models}
\begin{figure}[!ht]
\centering
\begin{subfigure}{0.48\textwidth}
\includegraphics[width=\linewidth]{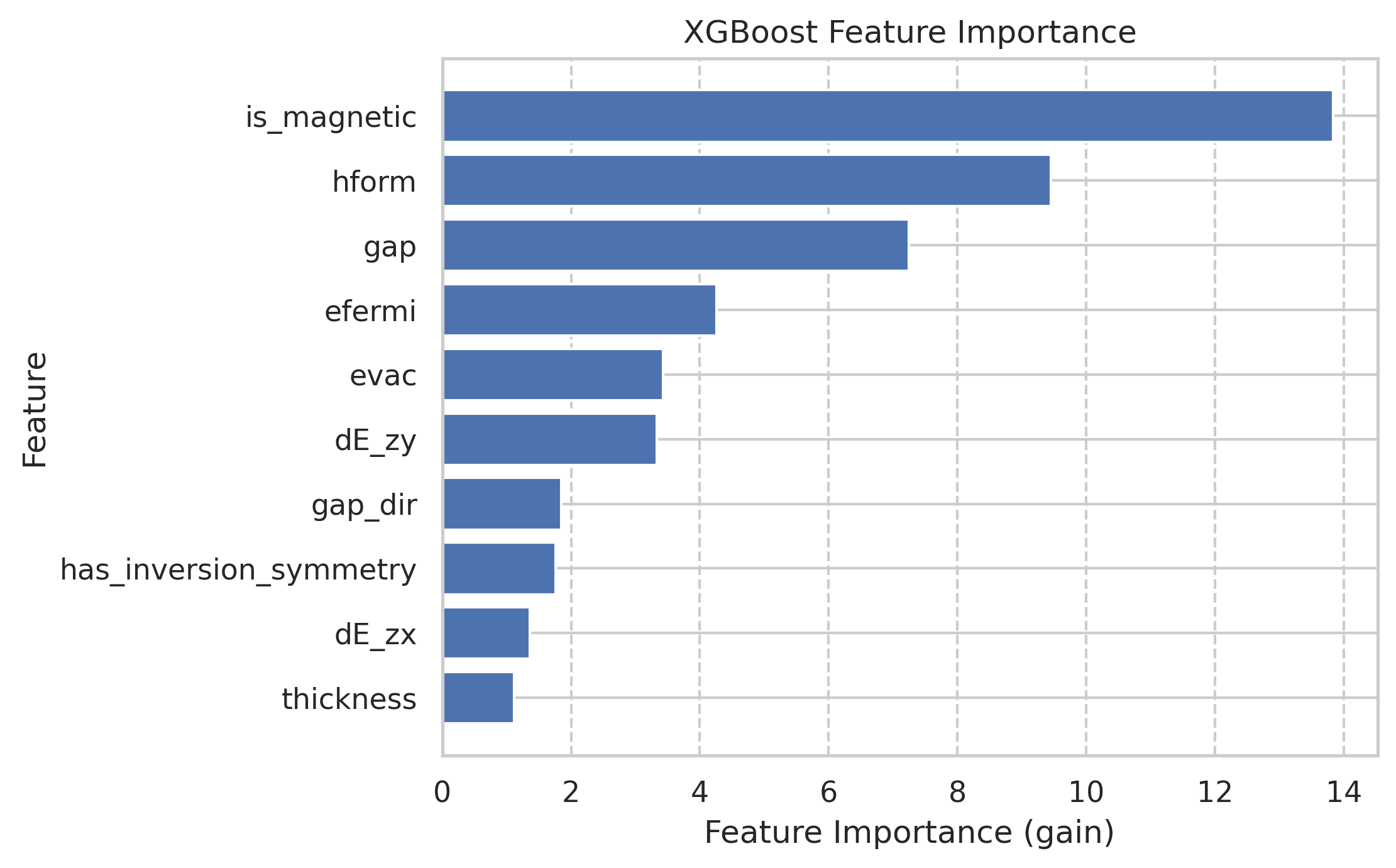}
\caption{C2DB native (top 10 of 19)}
\end{subfigure}
\begin{subfigure}{0.48\textwidth}
\includegraphics[width=\linewidth]{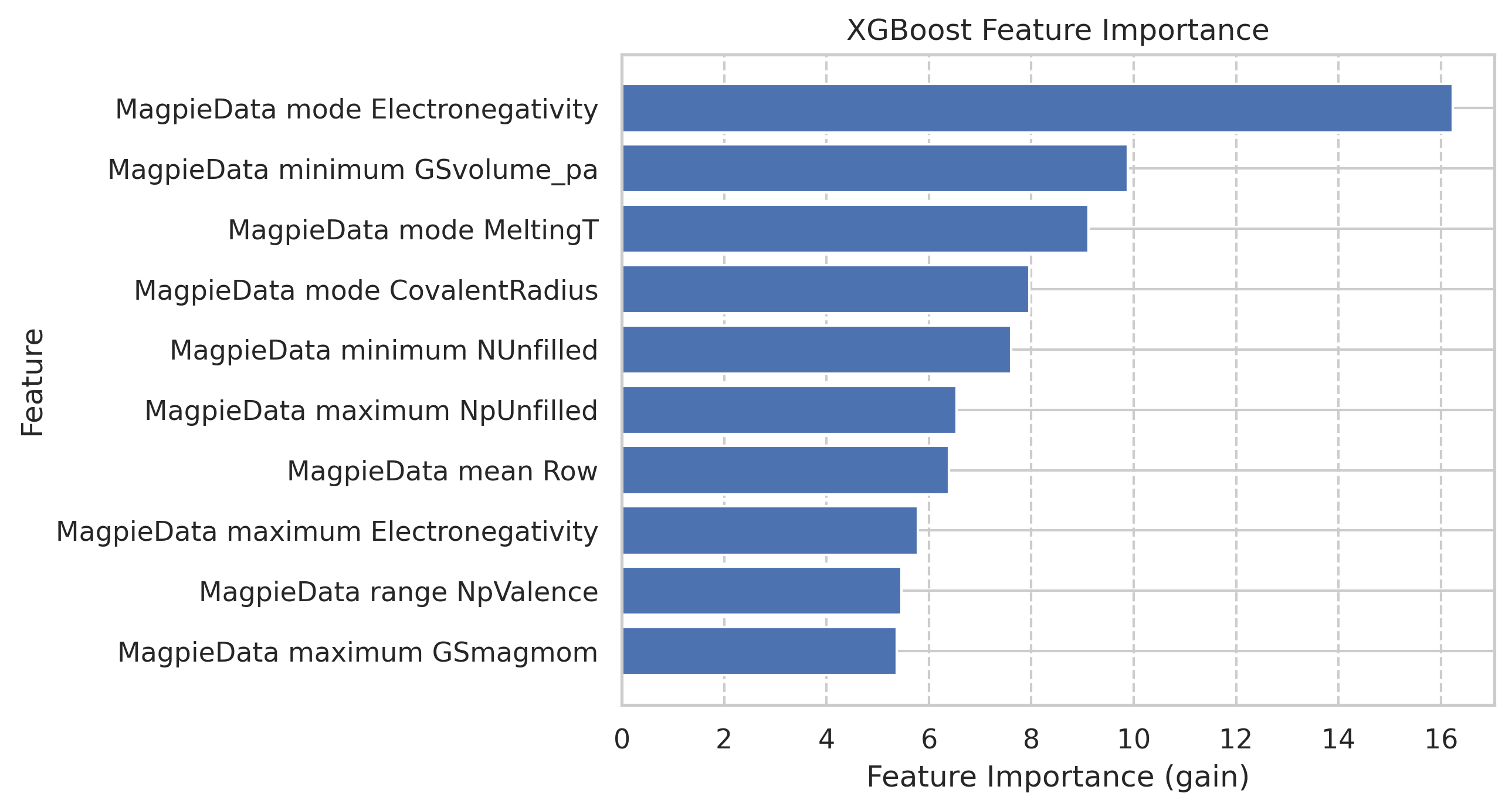}
\caption{Magpie + structural (top 10 of 141)}
\end{subfigure}
\caption{Gain-based XGBoost feature importance for the C2DB-native and Magpie+structural $\Delta$-learning regressors. The curated-model importance is shown in the main text.}
\label{figS:regimp}
\end{figure}

The main text reports the gain-based XGBoost feature-importance profile for the curated $\Delta$-learning regressor. Figure~\ref{figS:regimp} gives the corresponding C2DB-native and Magpie+structural profiles. In the 19-feature C2DB-native model, \feat{is\_magnetic} has more than four times the gain of the next-ranked descriptor, \feat{dE\_zy}, while the raw PBE gap ranks sixth among the displayed features. Thus, within this compact representation, the learned correction is not controlled by the PBE gap magnitude alone; magnetic and structural/energetic descriptors also carry substantial predictive signal.

The Magpie model is led by mode electronegativity and minimum ground-state volume per atom, the same two descriptors that lead the curated model. Their persistence after additional C2DB and local-environment features are added is consistent with the improvement of the curated model arising from complementary information rather than replacement of the composition signal. As noted in the main text, gain importance is not a causal attribution and can be redistributed among correlated descriptors.

\subsection{Absolute error versus PBE gap}
\begin{figure}[!ht]
\centering
\begin{subfigure}{0.32\textwidth}
\includegraphics[width=\linewidth]{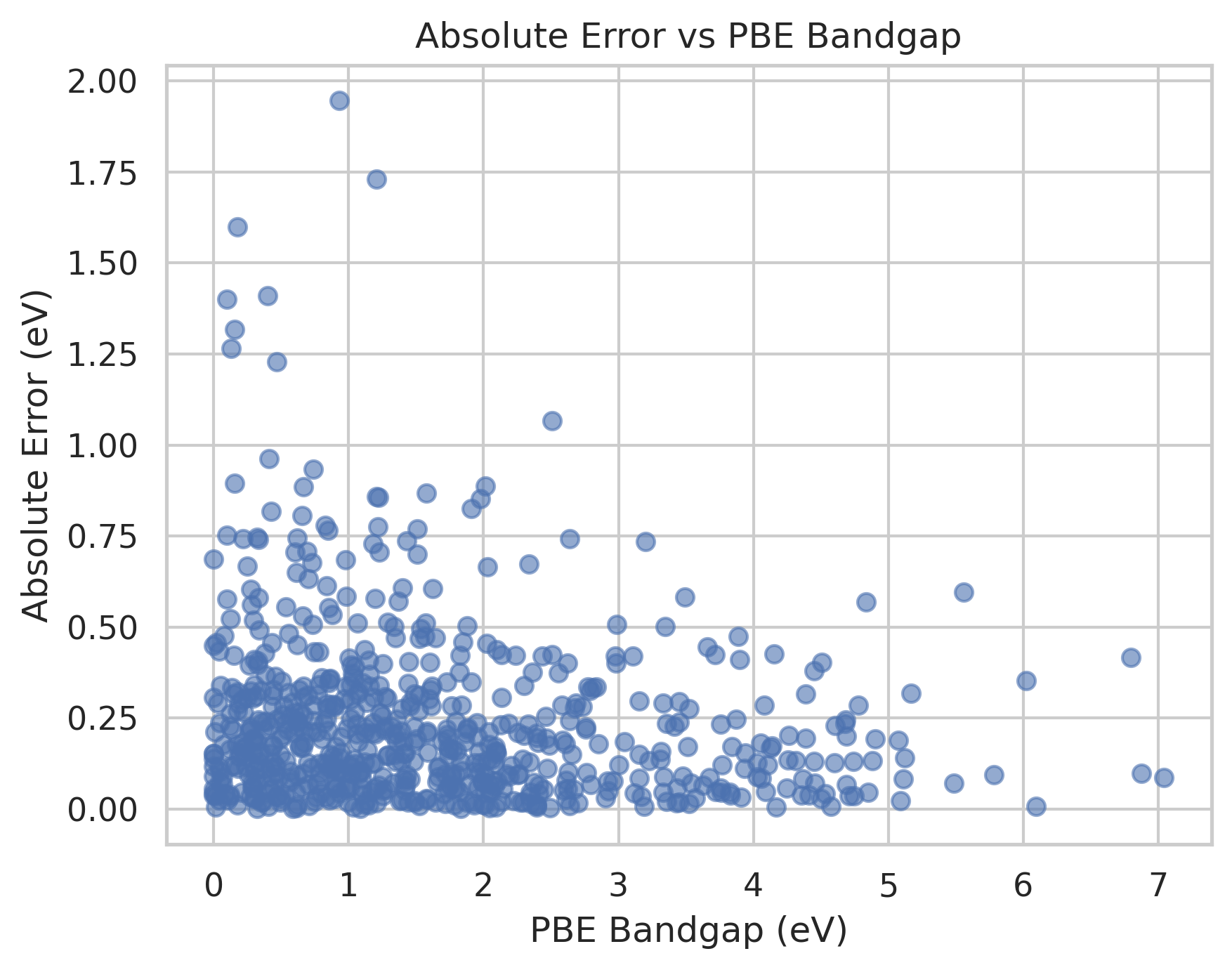}
\caption{C2DB native}
\end{subfigure}
\begin{subfigure}{0.32\textwidth}
\includegraphics[width=\linewidth]{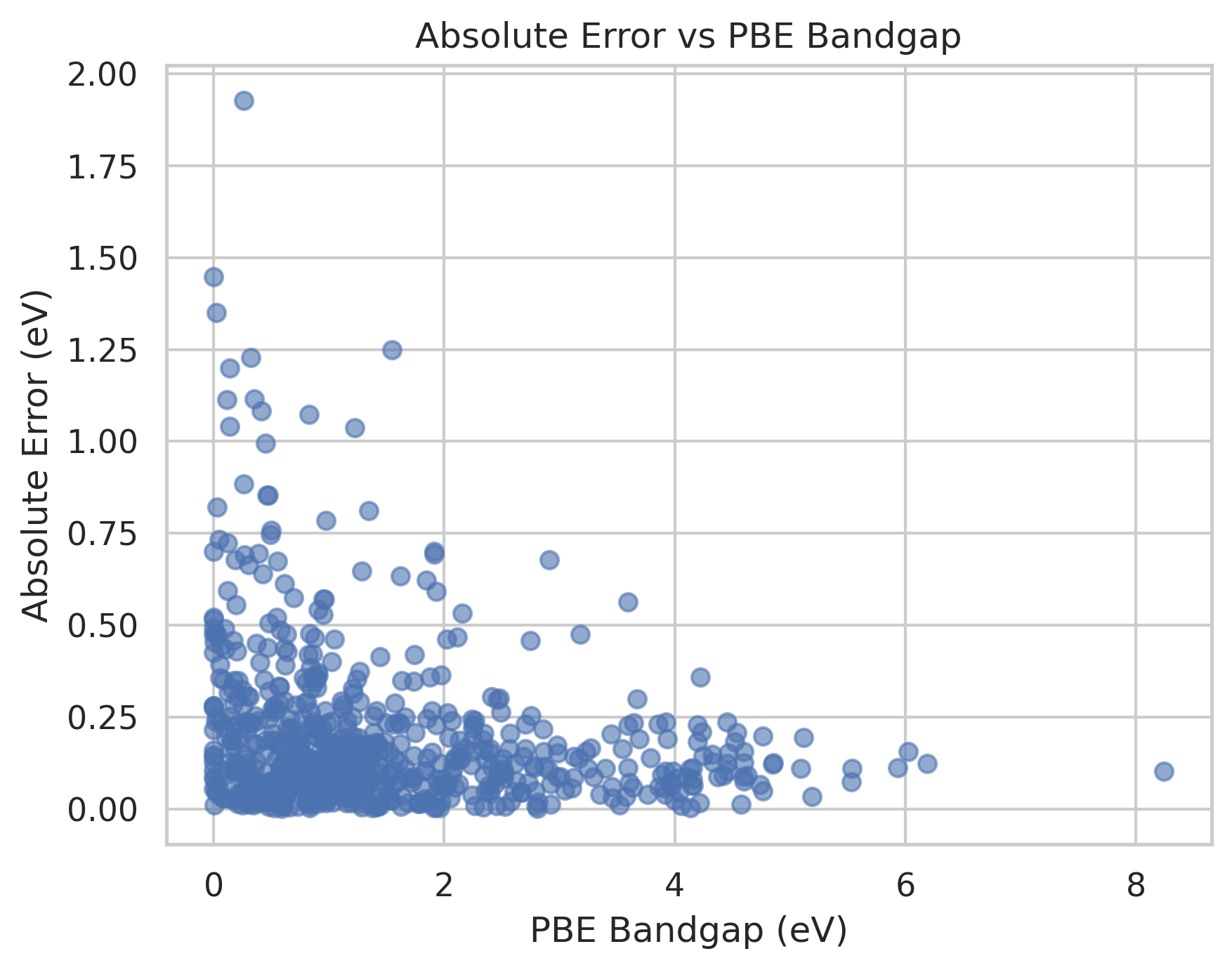}
\caption{Magpie + structural}
\end{subfigure}
\begin{subfigure}{0.32\textwidth}
\includegraphics[width=\linewidth]{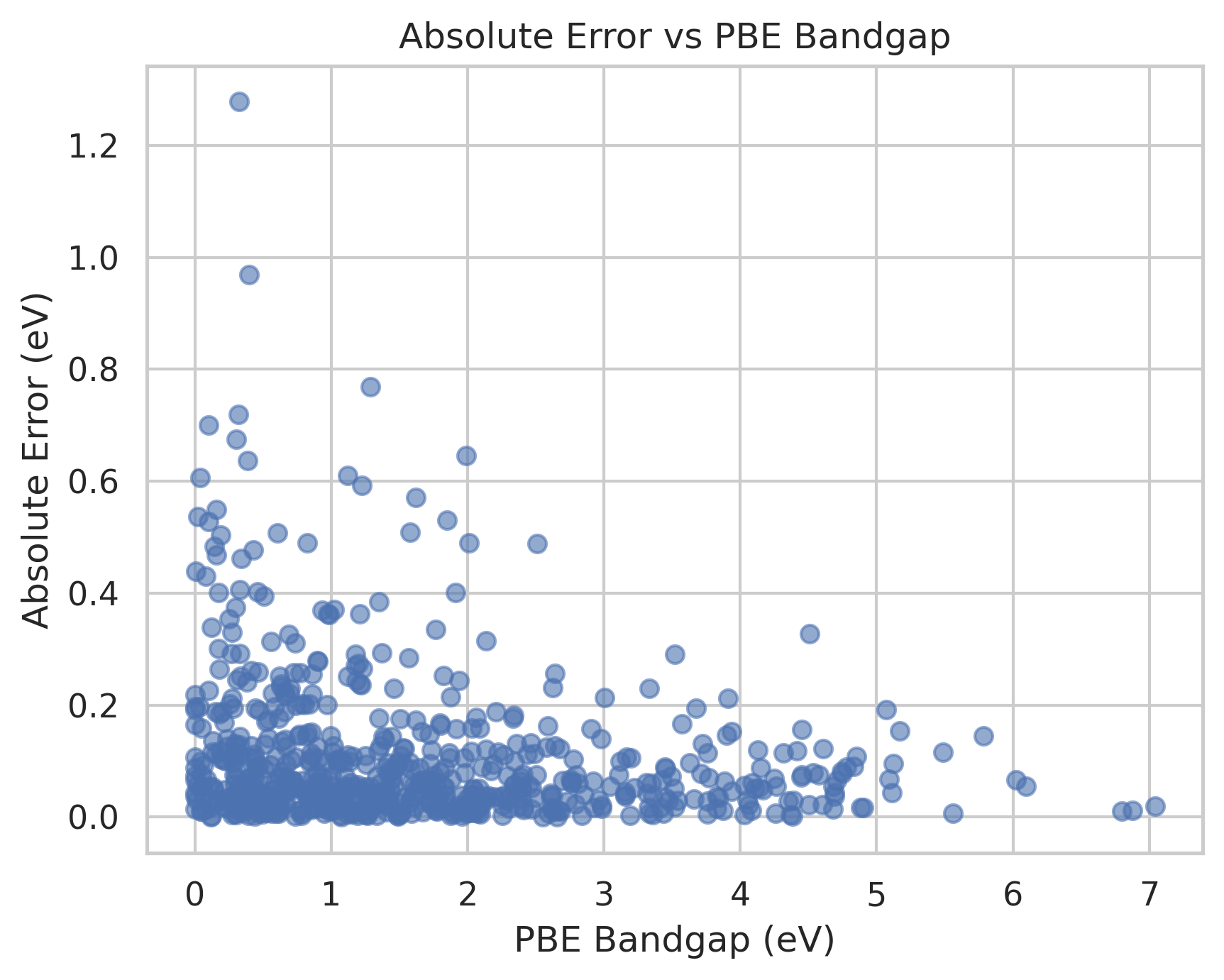}
\caption{Curated combined}
\end{subfigure}
\caption{Absolute HSE06-gap reconstruction error versus PBE gap on the 664-material regression test set.}
\label{figS:abserr}
\end{figure}

The main manuscript plots the signed $\Delta$-learning residual against the PBE gap. Figure~\ref{figS:abserr} shows the same test points as absolute error, making the heteroscedastic error envelope easier to read. For all three representations, the largest and most variable errors cluster at $\EgPBE<2$~eV and the envelope contracts strongly with increasing PBE gap. The curated model has the narrowest envelope across most of the gap range, consistent with its 0.108~eV overall MAE. The low-gap regime is therefore the most important target for uncertainty calibration and explicit HSE06 validation when a screening decision lies close to a numerical threshold.

\FloatBarrier

\section{Targeted HSE06 Validation of ML-Predicted Materials}
Table~\ref{tabS:spotvalidation} collects all six compounds for which the current calculation-status sheet contains both a completed explicit HSE06 calculation and ML predictions. Retaining the complete available set avoids selecting examples based on agreement with the model. The recalculated PBE gaps agree with C2DB to within 0.05~eV for every material (mean absolute difference 0.023~eV). Against the new HSE06 references, the C2DB PBE and recalculated PBE baselines have MAEs of 1.503 and 1.480~eV, respectively, whereas the C2DB-listed HSE-equivalent values give an MAE of 0.527~eV. The C2DB-native, Magpie+structural, and curated ML predictions give MAEs of 0.417, 0.182, and 0.110~eV, respectively.

\begin{table}[!ht]
\scriptsize
\centering
\caption{Detailed material-resolved HSE06 validation set. Values in parentheses are absolute errors relative to the new HSE06 reference. ``C2DB-listed HSE-equivalent'' reproduces the ML-tagged value in the calculation-status sheet and is not the same quantity as the C2DB-native descriptor model used in the manuscript.}
\label{tabS:spotvalidation}
\resizebox{\textwidth}{!}{%
\begin{tabular}{lccccccc}
\toprule
Material & HSE06 this work & PBE C2DB & PBE this work & C2DB-listed HSE-equivalent & C2DB-native ML & Magpie ML & Curated ML \\
 & (eV) & (eV; $|\epsilon|$) & (eV; $|\epsilon|$) & (eV; $|\epsilon|$) & (eV; $|\epsilon|$) & (eV; $|\epsilon|$) & (eV; $|\epsilon|$) \\
\midrule
\feat{1AgBr-1} & 2.97 & 1.57 (1.40) & 1.62 (1.35) & 2.49 (0.48) & 2.65 (0.32) & 2.70 (0.27) & 2.78 (0.19) \\
\feat{1AgCl-1} & 3.16 & 1.66 (1.50) & 1.68 (1.48) & 2.55 (0.61) & 2.75 (0.41) & 2.85 (0.31) & 3.03 (0.13) \\
\feat{1AgI-1}  & 2.70 & 1.51 (1.19) & 1.55 (1.15) & 2.40 (0.30) & 2.58 (0.12) & 2.60 (0.10) & 2.55 (0.15) \\
\feat{1BrCu-1} & 2.72 & 1.02 (1.70) & 1.03 (1.69) & 2.17 (0.55) & 2.14 (0.58) & 2.83 (0.11) & 2.74 (0.02) \\
\feat{1ClCu-1} & 2.84 & 1.09 (1.75) & 1.09 (1.75) & 2.19 (0.65) & 2.21 (0.63) & 3.00 (0.16) & 2.85 (0.01) \\
\feat{1CuI-1}  & 2.75 & 1.27 (1.48) & 1.29 (1.46) & 2.18 (0.57) & 2.31 (0.44) & 2.89 (0.14) & 2.91 (0.16) \\
\midrule
MAE & --- & 1.503 & 1.480 & 0.527 & 0.417 & 0.182 & 0.110 \\
\bottomrule
\end{tabular}%
}
\end{table}

\end{document}